\documentclass[review,12pt]{elsarticle}

\usepackage{amssymb}

\usepackage{amsmath} 
\usepackage{amsfonts} 

\usepackage{xcolor} 
\usepackage{graphicx}

\usepackage{tabularx}   
\usepackage{makecell}   

\usepackage{float}  

\usepackage{subfigure}  
\usepackage{multirow}

\usepackage{ulem}  

\biboptions{comma,sort&compress}

\usepackage[colorlinks=true, linkcolor=blue, citecolor=blue, urlcolor=blue]{hyperref}

\journal{Elsevier}

\begin{document}

\begin{frontmatter}



\title{Matrix representation and GPU-optimized parallel B-spline computing}

\author[]{Jiayu Wu}
\author[]{Qiang Zou\corref{cor}}\ead{qiangzou@cad.zju.edu.cn}

\cortext[cor]{Corresponding author.}
\address{State Key Laboratory of CAD$\&$CG, Zhejiang University, Hangzhou, 310058, China}

\begin{abstract}
B-spline modeling is fundamental to CAD systems, and its evaluation and manipulation algorithms currently in use were developed decades ago, specifically for CPU architectures. While remaining effective for many applications, these algorithms become increasingly inadequate as CAD models grow more complex, such as large-scale assemblies and microstructures. GPU acceleration offers a promising solution, but most existing GPU B-spline algorithms simply adapt CPU counterparts without accounting for the mismatch between the unstructured, recursive nature of B-splines and the structured nature of GPU kernels, ultimately failing to fully leverage GPU capabilities. This paper presents a novel approach that transforms B-spline representations into regular matrix structures, reducing all evaluation and manipulation computations to matrix addition and multiplication, thus better aligning with GPU architecture. By combining this matrix representation with GPU-optimized task scheduling and memory access patterns, the paper demonstrates significant performance improvements in the key B-spline operations of inversion and projection. Experimental results show an improvement of about two orders of magnitude in computational speed compared to existing methods.
\end{abstract}

\begin{keyword}
Computer-Aided Design \sep B-spline Curves and Surfaces \sep Matrix Representation \sep GPU Computing \sep Inversion and Projection
\end{keyword}

\end{frontmatter}


\section{Introduction}
\label{sec:sample1}
Computer-aided design (CAD) plays a crucial role in modern engineering and manufacturing~\cite{zou2024intelligent}, and B-splines serve as a fundamental method for representing the geometry of engineering products in CAD~\cite{piegl2012nurbs}; for instance, B-spline curves/surfaces are essential tools in shape design~\cite{zou2022robust}, feature modeling~\cite{zou2020decision}, engineering analysis~\cite{li2023xvoxel}, and process planning~\cite{zou2014iso}. However, the mathematical complexity and recursive computations involved in evaluating and manipulating B-splines often lead to performance bottlenecks, particularly when working with large-scale CAD models, such as complex assemblies or microstructures~\cite{zou2024geometric,huang2005interactive}. For instance, the slow computation of p-curves for B-spline surfaces in large CAD assemblies is a major cause of long model loading times~\cite{kovacs2017p}.

GPU parallel computing holds promise for addressing performance bottlenecks. While notable progress has been made in applying GPU acceleration to various computational tasks~\cite{guthe2005gpu,krishnamurthy2007direct,krishnamurthy2008performing,concheiro2010synthesis,ruijters2012gpu,cui2013real,palomar2018high}, many B-spline algorithms still rely on adaptations of CPU-based approaches. These adaptations often fail to account for the structural mismatch between B-splines' recursive nature and the highly parallelized architecture of GPUs, thus limiting the potential performance gains~\cite{zhou2010data}. A common example is the direct execution of de Boor’s recursion~\cite{piegl2012nurbs} on GPU kernels, which is not well-suited for the GPU’s parallel processing model.

Very recently, one emerging direction is the conversion of B-spline operations into matrix-based computations, which align more naturally with GPU's parallel processing capabilities~\cite{xiong2023eter}. However, while this approach has shown improvements for low-degree B-splines in B-spline evaluation tasks, it remains under-explored for higher-degree B-splines and more complex operations such as inversion, projection, and approximation.

This paper extends the concept of matrix-based GPU computation to high-degree B-splines and advanced operations. As mentioned earlier, fully leveraging GPU capabilities requires structured data and algorithms. Therefore, we address two key challenges that hinder GPU optimization: variation in B-spline degree and variation in knots and control points. We propose a decomposition method that transforms B-splines of arbitrary degree into a sequence of cubic Béziers, within a specified error threshold. This decomposition ensures uniformity in parameters across all Béziers, making them more suitable for GPU processing. While this technique is well-established in the CPU context~\cite{piegl2012nurbs}, we extend it by converting all related advanced operations---such as knot insertion~\cite{boehm1985efficiency}, degree elevation/reduction~\cite{piegl2012nurbs}, and approximation~\cite{zou2025splinegen}---into matrix representations and manipulations, enabling efficient GPU processing. An error-control mechanism is also introduced to manage approximation errors during this decomposition process, and once again, it is matrix-represented.

With this matrix representation (M-rep), B-splines are reformatted into a structure well-suited for GPU storage and manipulation. However, the decomposition alone does not guarantee optimal performance, as different B-spline curves are decomposed into varying-sized collections of Bézier curves. This can lead to unstructured task distribution, resulting in memory and warp divergences~\cite{rogers2013divergence, hong2011accelerating}, which undermine GPU parallelism. To address these issues, we propose optimization strategies for task scheduling, memory management, and workload balancing, ensuring efficient GPU resource utilization and maximizing parallelism.

To demonstrate the effectiveness of the above matrix-represented and GPU-optimized B-spline computing method, we focus on two fundamental operations---point inversion and projection---that highlight the potential of high-degree B-splines for efficient processing. These examples not only illustrate the feasibility of our method but also lay the foundation for applying it to more complex B-spline tasks in CAD systems.

The remainder of this paper is organized as follows: Sec.~\ref{sec:related_work} reviews the literature, and Sec.~\ref{sec:preliminaries} gives a very brief introduction to GPU computing. Sec.\ref{sec:problems} gives the problem statement. Sec.\ref{sec:Methods} elaborates the M-rep in detail. Validation of the method using a series of examples and comparisons can be found in Sec.~\ref{sec: results}, followed by conclusions in Sec.~\ref{sec:conclusion}.

\section{Related Work}
\label{sec:related_work}
In this section, we briefly discuss literature related to GPU B-spline computing (Sec.~\ref{sec:GPU for B-Spline computing}), GPU for projection and inversion (Sec.~\ref{sec:GPU for inversion/projection}), and GPU optimization methods (Sec.~\ref{sec:GPU optimized methods}). 

\subsection{GPU for B-spline computing}
\label{sec:GPU for B-Spline computing}
Existing GPU B-spline computing methods can be divided into two categories: direct methods and decomposition-based methods.

\textbf{Direct methods.}
Direct methods operate directly on B-spline representations, including power basis~\cite{gatilov2016vectorizing} and Bernstein basis~\cite{krishnamurthy2007direct,concheiro2014interactive,zhao2024tpms2step}, eliminating the need for additional storage and reducing data redundancy. However, for high-order B-splines, these methods involve multiple control points and complex weight calculations, leading to strong data dependencies that make efficient GPU parallelization challenging.

\textbf{Decomposition-based methods.}
Decomposition-based methods involve breaking down complex B-splines into simpler components such as lower-order curves and surfaces~\cite{guthe2005gpu} or piecewise Béziers~\cite{xiong2023eter} before applying other operations. These techniques can enhance GPU efficiency by enabling independent computation of each component. However, the additional decomposition steps may introduce potential errors.

\subsection{GPU for projection/inversion}
\label{sec:GPU for inversion/projection}
Using GPUs to accelerate point inversion and projection onto curves and surfaces is an intuitive approach, but research in this area is limited. Krishnamurthy et al. \cite{krishnamurthy2008performing} proposed a GPU-based method for point inversion on B-spline surfaces. By recursively subdividing the surface, constructing bounding boxes, and identifying the bounding box that contains the target point, this method can quickly find the inversion of a target point.

Although this algorithm leverages GPU parallelism, most bounding boxes do not actually contain the target point. As a result, constructing a large number of bounding boxes introduces significant computational overhead, which further degrades performance when handling large-scale tasks.

\subsection{GPU optimization methods}
\label{sec:GPU optimized methods}
GPU optimization methods are software or hardware techniques proposed to improve GPU application performance~\cite{hijma2023optimization}. Here we provide a summary of three software approaches.

\textbf{Kernel fission.}
While GPUs have a highly regular architecture, many applications and algorithms exhibit irregularities that hinder efficient execution. To better map irregular algorithms to GPUs, kernel fission has been proposed, which involves dividing a single kernel into multiple kernels or splitting a single kernel iteration into multiple iterations \cite{ashari2014fast, dalton2015optimizing, abdelfattah2013systematic, bos2010performance}.

\textbf{Coalesced memory access.}
Coalesced memory access improves memory efficiency by ensuring that a warp’s memory requests satisfy coalescing rules, reducing the number of transactions needed to fetch data from device memory~\cite{zou2024meta}. Various techniques have been proposed to achieve this, including reorganizing thread access patterns \cite{delbosc2014optimized, hu2018tricore} and selecting an optimal thread block size \cite{leist2009exploiting}.

\textbf{Workload balancing.}
Workload imbalance occurs when some threads remain idle while others perform useful work. To make workload balanced, techniques such as deferring outliers using a global work list \cite{hong2011accelerating}, distributing workloads within thread blocks or warps \cite{10.1145/2145816.2145832, wang2016gunrock, yang2018design}, and partitioning-based grouping strategies \cite{davidson2014work, wang2016gunrock} have been proposed to improve efficiency.

\section{GPU Preliminaries}
\label{sec:preliminaries}

\subsection{GPU architectures}
\begin{figure}[b]
    \centering
    \includegraphics[width=1\linewidth]{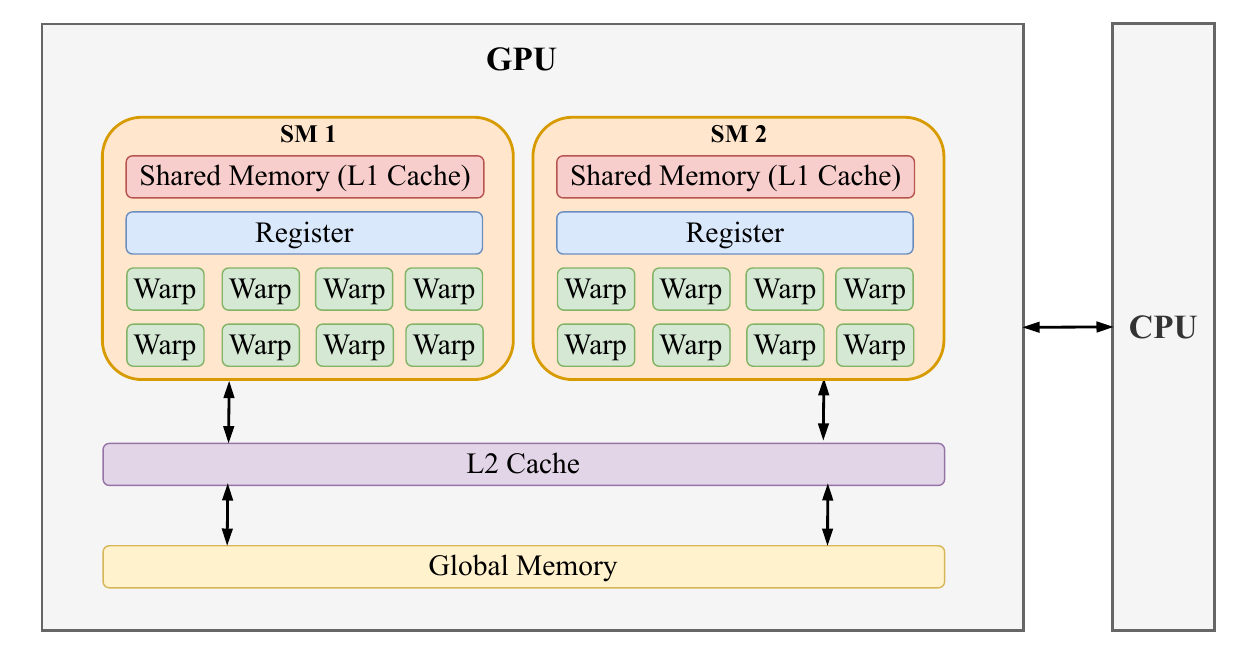}
    \caption{An overview of GPU architecture.}
    \label{fig:GPU architecture}
\end{figure}

GPUs consist of several streaming multiprocessors (SMs). As shown in Fig.~\ref{fig:GPU architecture}, each SM has its own local resources, such as shared memory (L1 cache), registers, and several warps, where each warp is a group of threads. SMs share global resources, including global memory, L2 cache, which is positioned between the L1 cache and global memory, etc.
Data is transferred between global memory and caches in units called `cache line', typically 128 bytes, determined by GPU hardware.

In practice, GPUs divide computational resources into warps, each containing multiple threads (e.g., 32). Threads within a warp execute the same instruction in parallel. Warp divergence refers to the problem of threads in a warp taking different branches, which have to be executed in serial,  resulting in reduced parallelism.
If memory transactions in a warp cannot be coalesced into a single cache line, multiple memory transactions are required, leading to inefficiency as threads must wait for data.
This is a phenomenon known as memory divergence.

To fully leverage the parallel capabilities of the GPU, an application should focus on the following aspects:
\begin{itemize}
    \item \textbf{Warp-centric programming.} 
    Since GPUs employ warps as the smallest scheduling unit, it is essential to integrate them into the programming model by structuring code around warps as fundamental units of computation.
    \item \textbf{Memory and warp divergence reduction.} Both types of divergence lead to idle GPU time and underutilized cores, reducing overall performance. Coalescing memory transactions and kernel fission are keys to reducing divergence.
\end{itemize}

\subsection{The recent tensor core techniques}
Tensor cores~\cite{markidis2018nvidia} are specialized hardware units in modern GPUs, designed specifically to accelerate matrix operations. 
A single tensor core can process multiple elements in parallel, drastically increasing computational throughput.

However, effectively utilizing tensor cores is not trivial. Their design is optimized for structured matrix multiplication and addition, requiring computations to be formulated in a compatible matrix format. This constraint makes it difficult to directly apply tensor cores to unstructured problems such as B-spline computations, which typically involve irregular data structures and iterative processes. Therefore, transforming these computations into suitable matrix operations is critical to fully leverage the performance benefits of tensor cores.

\section{Problem Statement}
\label{sec:problems}
Mathematically, a B-spline curve of degree $p$ with control points $\overline{\mathbf{P}} = \{ P_i \in \mathbb{R}^3 \mid i = 0, 1, \ldots, m\}$ and knot vector $\overline{\mathbf{T}} = \{ t_0, t_1, \cdots, t_{m+p+1} \}$ is defined as:
\begin{equation}
    C(t) = \sum_{i=0}^{m} {N_{i,p}}(t) {P_i}  \quad  t \in [t_0, t_{m+p+1}]\   
\end{equation}
where ${N_{i,p}}(t),\ i=0,\dots,m,$ are the B-spline basis functions of degree $p$. It is evaluated as:
\begin{equation}
\label{eq:bsplinebasis}  
    \begin{aligned}
        &N_{i,0}(t)= 
            \begin{cases}
                1,\quad &t_i\leq{t}<t_{i+1} \\
                0,\quad &otherwise
            \end{cases} \\
        &N_{i,p}(t)=\frac{t-t_i}{t_{i+p}-t_i}N_{i,p-1}(t)+\frac{t_{i+p+1}-t}{t_{i+p+1}-t_{i+1}}N_{i+1,p-1}(t)     
    \end{aligned} 
\end{equation}

Given a point $\mathbf{q}\in \mathbb{R}^3$, 
the problem of projecting $\mathbf{q}$ onto $C(t)$ can be described mathematically as to find $t^*$ ,such that:
\begin{equation}
\|\mathbf{q} - C(t^*)\| = \min \{\|\mathbf{q} - C(t)\| \ | \ t \in [t_0, t_{m+p+1}]\}
\label{eq:projection}
\end{equation}

For a set of test points $\overline{\mathbf{Q}} = \{\mathbf{q}_{i} \in \mathbb{R}^3 \ | \ i = 0, 1, \ldots, N\}$, our task is to find the projection parameter set \( S \), which is defined as:
\begin{equation}
S = \{t_i^* \ | \ t_i^* = \arg\min_{t \in [t_0, t_{m+p+1}]} \|\mathbf{q}_{i} - C(t)\|, \ i = 0, 1, \ldots, N\}
\label{eq:projections}
\end{equation}
where \( t_i^* \) is the parameter for the closest point on \( C(t) \) to \( q_i \). 

When a single test point \( \mathbf{q}_{i} \) lies on the curve \( C(t) \), the projection problem degenerates into the inversion problem. At this time, $t_i^*$ can be written as:
\begin{equation}
t_i^* = C^{-1}(\mathbf{q}_{i}), \quad t_i^* \in [t_0, t_{m+p+1}]
\label{inversion}
\end{equation}
where \( C^{-1} \) is the inverse function of the curve \( C(t) \).

\section{Methods}
\label{sec:Methods}
\subsection{Methods overview}
\begin{figure*}
    \centering
    \includegraphics[width=1.0\linewidth]{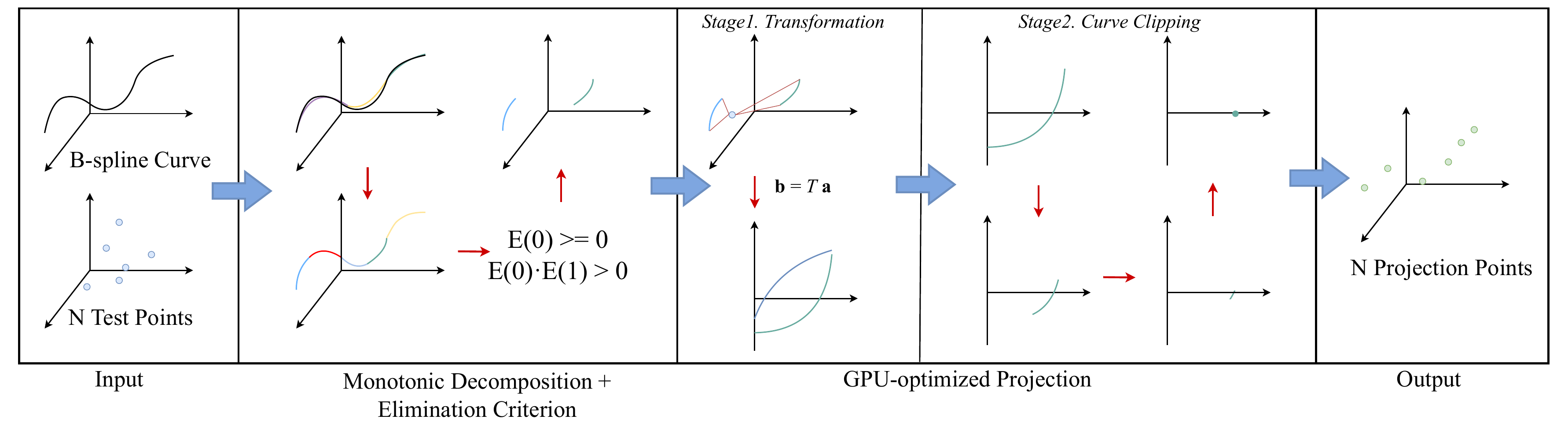}
    \caption{The pipeline of M-rep.}
    \label{fig:total_pipeline}
\end{figure*}

Since Béziers have a simpler mathematical formulation compared to B-splines, they are more suitable for efficient parallel execution on GPUs using matrix operations.
We first reformulate the decomposition of B-spline curves into piecewise Béziers as a series of matrix operations suitable for tensor cores (Sec.~\ref{M-rep for the decomposition}).
To reduce the computational complexity of higher-order Béziers, a matrix-based approximation method that approximates them with cubic Béziers while maintaining a controllable tolerance is proposed (Sec.~\ref{error-controlled decomposition}). The choice of cubic Béziers is driven by two factors.  
First, the polynomial coefficient matrix of the cubic Bézier curve has 16 elements, which makes it easy to meet the tensor core memory alignment requirements~\cite{nvidia2021cublas}. Second, the Gram matrix~\cite{rababah2003distance} $G_{3,3}$ (Sec.~\ref{error-controlled decomposition}) associated with two cubic Bézier functions is $4 \times 4$, and when using double-precision data types, it fully occupies a 128-byte cache line on the GPU, optimizing memory access efficiency. In addition, a GPU scheduling algorithm is designed to accelerate our approximation process (Sec.~\ref{error-controlled decomposition}).

For point projection and inversion on B-spline curves, a globally optimal method is employed to subdivide cubic Bézier curves into monotonic segments, ensuring that each segment contains at most one potential projection point (Sec.~\ref{monotonic}). 
After obtaining the monotonic segments, we introduce an elimination criterion to filter out segments that do not contain potential projection points (Sec.~\ref{M-rep for bezier inversion/projection}). 
For the remaining segments, the derivatives of the squared distance functions from points to these segments are converted into non-parametric Bézier representations, thus we utilize Bézier clipping~\cite{sederberg1990curve} to efficiently locate the projection parameters (Sec.~\ref{M-rep for bezier inversion/projection}).
The monotonicity property ensures that no further subdivision is required in Bézier clipping, transforming the complex problem of solving nonlinear equations into a simple line and line-segment intersection problem, thereby significantly reducing GPU overhead. This transformation is also formulated as a matrix operation, enabling efficient computation on tensor cores and significantly accelerating the conversion process (Sec.~\ref{M-rep for bezier inversion/projection}). 

To prevent workload imbalance, we design a workload scheduled algorithm for Bézier clipping that adjusts the number of intersections each thread computes (Sec.~\ref{sec:warp-centric projection/inversion}).
Additionally, a point cooperative work sharing strategy is proposed to ensure the correctness of projection point updates while minimizing the impact on parallel execution speed (Sec.~\ref{PCWS}). Our general process is illustrated in Fig.~\ref{fig:total_pipeline}.

\subsection{B-spline to Bézier decomposition}
\label{ECMCB(Error-controlled Monotonic Cubic Bezier) approximation}
To enable efficient matrix operations on GPUs, we first introduce a method to convert B-splines into piecewise Béziers using tensor cores, and then cubic Béziers are used to approximate the piecewise Béziers under controllable error. Finally, for the projection and inversion problems, a novel decomposition strategy is specially designed to find the global optimal solution.

\subsubsection{M-rep for the decomposition}
\label{M-rep for the decomposition}
The traditional approach to decomposing a B-spline into piecewise Béziers through knot insertion is inherently a serial process~\cite{piegl2012nurbs}, making it inefficient when directly ported to the GPU.
This limitation becomes more pronounced as the number of knots increases.
To address it, we redesign the decomposition process by formulating it as a series of structured matrix operations.
Through this transformation, we identify computational invariants within each knot interval and extract them to eliminate redundant calculations. Additionally, the presence of these invariants allows us to leverage tensor cores to further accelerate the entire decomposition process.

\begin{figure}
    \centering
    \includegraphics[width=0.9\linewidth]{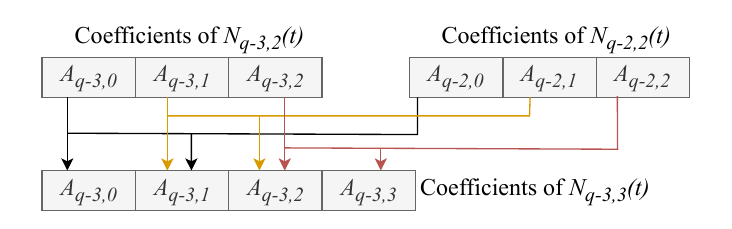}
    \caption{The workflow of vectorizing the polynomial coefficients of ${N_{q-3,3}}(t)$ in $A_{q,p,T}$ within the interval \( t \in [t_q, t_{q+1}] \).}
\label{fig:pre_compute}
\end{figure}

Since a $p$-degree B-spline curve $C(t)$ is locally determined by $p+1$ control points,
its expression within the interval \( t \in [t_q, t_{q+1}] \) can be simplified as:
\begin{equation}
C(t) = \sum_{i=q-p}^{q} N_{i,p}(t) P_i
\label{eq:bsplineinterval}
\end{equation}
which can be further rewritten in matrix form as:
\begin{equation}
C(t) = \begin{bmatrix} 1 & t & t^2 & \cdots & t^{p} \end{bmatrix} 
A_{q,p,T} 
\begin{bmatrix} 
P_{q-p} \\
P_{q-p+1} \\
\vdots \\
P_q
\end{bmatrix}
\label{eq:bsplinematrix}
\end{equation}
where $A_{q,p,T}$ denotes the matrix of polynomial coefficients associated with the basis functions ranging from $N_{q-p,p}(t)$ to $N_{q,p}(t)$. After decomposition, Eq.~\eqref{eq:bsplineinterval} can be expressed as: 
\begin{equation}
C(t) = R(u) = \sum_{i=0}^{p} B_{i,p}(u) Q_{q,i} \quad  u \in [0, 1]\
\label{eq:bezierinterval}
\end{equation}
where $B_{i,p}(u)$ represents the Bernstein basis function, $Q_{q,i}$ refers to the $i$-th control point of the Bezier segment corresponding to the interval \( t \in [t_q, t_{q+1}] \). Similarly, we can rewrite Eq.~\eqref{eq:bezierinterval} in matrix representation as:
\begin{equation}
C(t) = R(u) = \begin{bmatrix} 1 & u & u^2 & \cdots & u^{p} \end{bmatrix} 
B_{p} 
\begin{bmatrix} 
Q_{q,0} \\
Q_{q,1} \\
\vdots \\
Q_{q,p}
\end{bmatrix}
\label{eq:bsplineafter}
\end{equation}
where $B_{p}$ is a matrix that encodes the Bernstein basis coefficients for degree $p$, which is constant.
Because the polynomial coefficients in $N_{i,j}(t)$ remain constant within the interval \( t \in [t_q, t_{q+1}] \), we can vectorize the computation of B-spline basis functions on the GPU, enabling their efficient parallel evaluation and storage in $A_{q,p,T}$, as illustrated in Fig.\ref{fig:pre_compute}.

After getting $A_{q,p,T}$, a reparameterization is required. The parameter range \( t \in [t_q, t_{q+1}] \) normalized to a local parameter \( u \in [0, 1] \) is as follows:
\[
u = \frac{t - t_q}{t_{q+1} - t_q}, \quad t = u (t_{q+1} - t_q) + t_q.
\]
the transformation matrix \( M_{q,p,T}  \) is defined such that:
\begin{equation}
T = U M_{q,p,T} 
\label{eq:Mqpt}
\end{equation}
where \( T = [1, t, t^2, \dots, t^p] \), \( U = [1, u, u^2, \dots, u^p] \).
By solving Eq.~\eqref{eq:bsplinematrix} and Eq.~\eqref{eq:bsplineafter} we can get
\begin{equation}
\begin{bmatrix} 
Q_{q,0} \\
Q_{q,1} \\
\vdots \\
Q_{q,p}
\end{bmatrix} = B_{p}^{-1} M_{q,p,T} 
A_{q,p,T} 
\begin{bmatrix} 
P_{q-p+1} \\
P_{q-p+2} \\
\vdots \\
P_q
\end{bmatrix}
\label{eq:decomposition}
\end{equation}
define :
\begin{equation}
D_{q,p} = M_{q,p,T} 
A_{q,p,T} 
\begin{bmatrix} 
P_{q-p+1} \\
P_{q-p+2} \\
\vdots \\
P_q
\end{bmatrix}
\label{Dqp}
\end{equation}
substituting Eq.~\eqref{Dqp} into Eq.~\eqref{eq:decomposition}, we finally get
\begin{equation}
\begin{bmatrix} 
Q_{q,0} \\
Q_{q,1} \\
\vdots \\
Q_{q,p}
\end{bmatrix} = B_{p}^{-1} D_{q,p}
\label{eq:decomfinal}
\end{equation}

\begin{figure}
    \centering
    \includegraphics[width=0.9\linewidth]{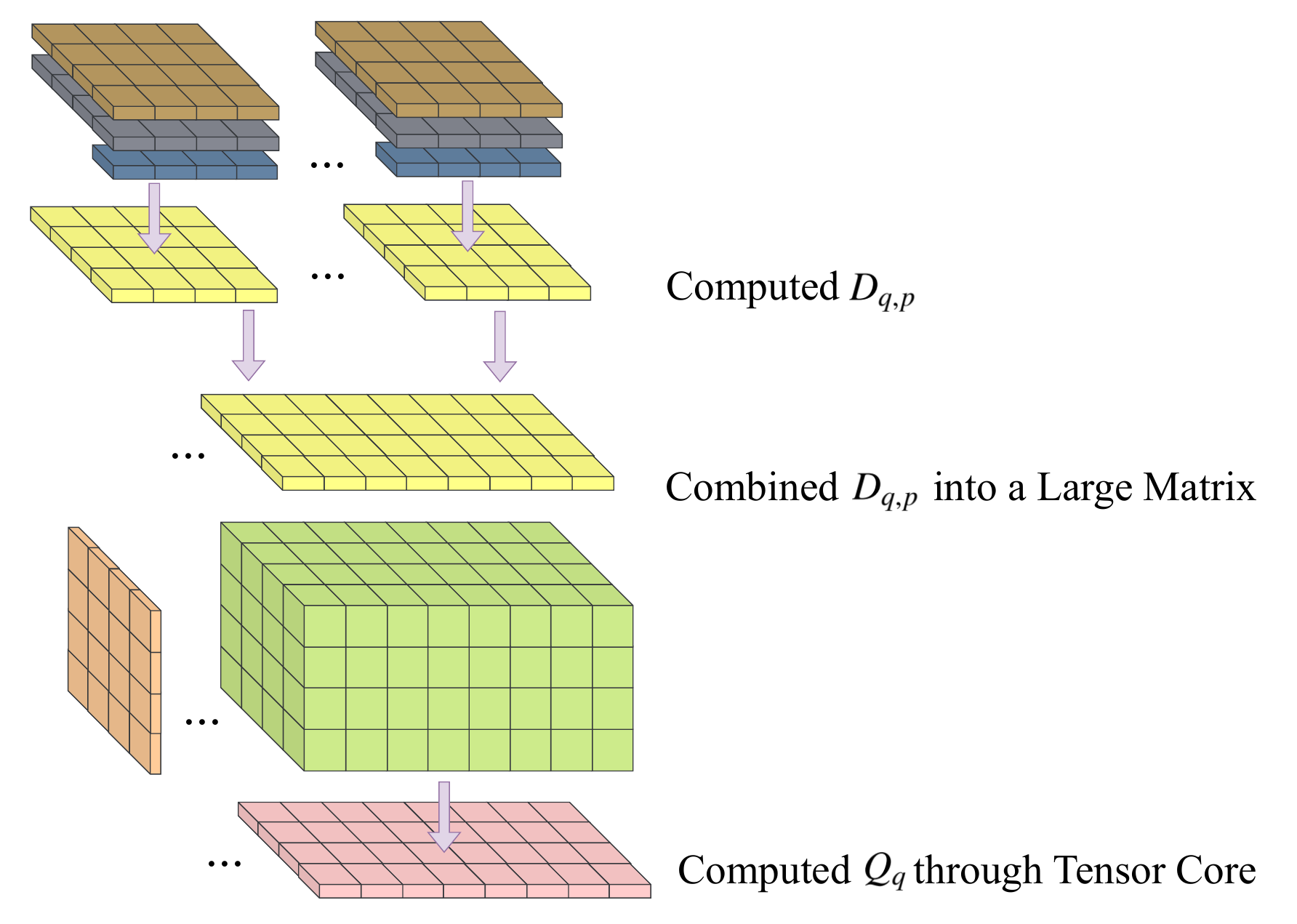}
    \caption{Pipeline of tensor core-accelerated B-spline decomposition.
    }
    \label{fig:decomposition}
\end{figure}

The computation of $D_{q,p}$ is carried out through multiple parallel small matrix multiplications. Since tensor cores are more efficient at performing two large matrix multiplications rather than processing multiple small matrices in parallel, an optimization of Eq.~\eqref{eq:decomfinal} is applied by leveraging computational invariants. Specifically, as each $D_{q,p}$ is multiplied by the same $B_{p}^{-1}$, we merge all $D_{q,p}$ into a single large matrix. This allows us to perform a single tensor core operation to multiply $B_{p}^{-1}$ with the entire large matrix, efficiently obtaining the control points for all piecewise Béziers in one step, as illustrated in Fig.~\ref{fig:decomposition}. We prove that reparametrizing $T$ is more efficient than reparametrizing $U$ in our decomposition process. Our proof and further derivation of the decomposition process are provided in supplementary materials.

\subsubsection{Error-controlled decomposition}
\label{error-controlled decomposition}

To address the complexity of handling high-degree B-spline data, the original piecewise Béziers are further decomposed into cubic Béziers while maintaining an $L_1$-error tolerance of $\alpha$. This decomposition process consists of two steps: first, the degree of the original Bézier curves is reduced. Then, based on these reduced Bézier curves, the original curves are approximated using progressively smaller segments.

\begin{figure}
    \centering
    \includegraphics[width=0.95\linewidth]{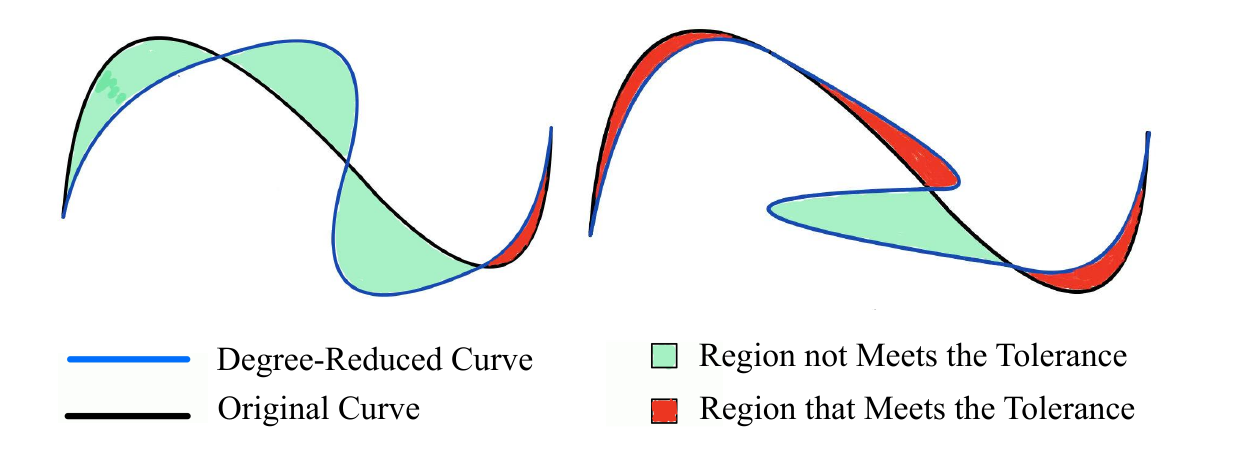}
    \caption{The degree-reduced curve after $L_1$-error minimization (left) and $L_2$-error minimization (right).}
    \label{fig:L1vsL2}
\end{figure}

In the initial degree reduction process, we ensure that the $L_2$-error $\epsilon$ is minimized, and the $G^1$-continuous condition is applied.
The choice of $L_2$-error minimization over $L_1$-error minimization during the first degree reduction is intentional, as $L_2$-error minimization ensures that the average $L_1$-error across the remaining segments is minimized. This approach facilitates the elimination of excessive segmentation in the subsequent approximation process, as illustrated in Fig.~\ref{fig:L1vsL2}.

Then a matrix-based approximation strategy is employed, incorporating subdivision and control polygon modification to ensure that the approximated cubic Bézier segments meet the specified tolerance requirements. Additionally, a GPU resource scheduling algorithm is designed to optimize the entire approximation process.

\textbf{$G^1$-continuous degree reduction.}
For the $q$-th piecewise Bezier curve of degree $p$ with control points $\mathbf{Q}=\begin{bmatrix}
Q_{q,0}, Q_{q,1} \dots Q_{q,p}
\end{bmatrix}^T$ obtained after decompostion: 
\[
Q(t) = \sum_{i=0}^{p} B_{i,p}(t) Q_{q,i} = B_{p} \mathbf{Q}, \quad t \in [0, 1],
\]
we need to find an approximated cubic Bézier curve:
\[
R(t) = \sum_{i=0}^{3} B_{i,3}(t) R_{q,i} = B_{3} \mathbf{R}, \quad t \in [0, 1], 
\]
with the control points $\mathbf{R}=\begin{bmatrix}
R_{q,0}, R_{q,1} \dots R_{q,3}
\end{bmatrix}^T$, so that the following two conditions are satisfied:
\begin{enumerate}
    \item[(1)] $Q(t)$ and $R(t)$ are $G^1$-continuous at the endpoints.
    \item[(2)] The $L_2$-error $\epsilon$ between $Q(t)$ and $R(t)$ is minimum.
\end{enumerate}
In this subsection, we use $Q_i$ to refer to $Q_{q,i}$, $R_i$ to refer to $R_{q,i}$ for convenience.

Using the definition of rababah~\cite{rababah2013linear}, we define matrix $G_{m,n}$ as a $(m+1)\times(n+1)$ matrix, whose elements are the integrals of products of Bernstein polynomials as follows:
\begin{equation}
g_{ij} = \int_{0}^{1} B_i^m(t) B_j^n(t) \, dt = \frac{\binom{m}{i} \binom{n}{j}}{(m+n+1) \binom{m+n}{i+j}}
\label{eq:gram}
\end{equation}

For $G^1$-continuous degree reduction constraint, the following conditions must be satisfied:
\begin{equation}
\begin{aligned}
R_{0} = Q_{0} &  
\quad R_{1} = Q_{0} + \frac{p}{3}  \Delta Q_{0}\delta_0 \\
R_{3} = Q_{p} & 
\quad R_{2} = Q_{p} - \frac{p}{3}  \Delta Q_{p-1}\delta_1
\end{aligned}
\label{G1-condition}
\end{equation}
where $\Delta Q_{i}$ means $Q_{i+1} - Q_{i}$, $\delta_0$ and $\delta_1$ are two variables.
And minimizing $L_2$-error means we need to minimize:
\begin{equation}
\varepsilon = \int_{0}^{1} \|B_p Q - B_3 R\|^2 \, dt
\label{epsl}
\end{equation}
the minimum occurs when the partial derivatives vanish. Differentiating it with respect to $\delta_i$ and equating to zero gives
\begin{equation}
\frac{\partial \varepsilon}{\partial \delta_0} = 
\left( G_{3,p}^1 Q - G_{3,3}^{1} R \right) \cdot \Delta Q_{0} = 0
\label{differentiatedelta0}\end{equation}
\begin{equation}
\frac{\partial \varepsilon}{\partial \delta_1} = 
\left( G_{3,p}^{2} Q - G_{3,3}^{2} R \right) \cdot \Delta Q_{p-1} = 0
\label{differentiatedelta1}
\end{equation}
where \( G_{m,n}^{i} \) is the submatrix of \( G_{m,n} \) formed by the $i$-th rows.

Note that Eq.~\eqref{differentiatedelta0} and Eq.~\eqref{differentiatedelta1} give a system of $2$ equations in $2$ unknowns, this system can be solved linearly. Our detailed derivation process is in supplementary materials. And because all \( G_{m,n} \) are invariant and can be pre-loaded into GPU memory without computing, the entire reduction process can be performed very quickly on the GPU.

\textbf{Error-controlled approximation.}
After obtaining all the cubic Bézier segments, we calculate the $L_1$-error for each segment and subdivide segments that do not meet the specified tolerance $\alpha$.
The subdivision process is performed via matrix operations on the GPU, and can be written as:
\begin{equation}
\mathbf{P} S(z) =
\begin{bmatrix}
\mathbf{P_{new}}(L) \\
\mathbf{P_{new}}(R) 
\end{bmatrix}
\label{submatrix}
\end{equation}
where $\mathbf{P}$
is the control points of the current cubic Bézier segment, $\mathbf{P_{new}}(L)$ is the control points of the left Bézier curve after subdivision, $\mathbf{P_{new}}(R)$ is the control points of the right Bézier curve after subdivision, and the subdivision matrix $S(z)$ is a matrix of $8 \times 4$, which means that the Bézier curve is divided into two parts at $t=z$. It can be written as:
\begin{equation}
S(z) =
\begin{bmatrix}
S_L(z) \\
S_R(z)
\end{bmatrix}
\label{subLandR}
\end{equation}
where $S_L(z)$ and $S_R(z)$ are both $4 \times 4$ matrices, representing the left and right subdivision matrix, respectively.
The nonzero elements of \( S_L(z) \) are:
\[
(S_L)_{i,j} = \binom{i}{j} z^j (1-z)^{i-j}, \quad 0 \leq j \leq i \leq 3,
\]
and the nonzero elements of \( S_R(z) \) are:
\[
(S_R)_{i,j} = \binom{3-i}{j} z^{j-i} (1-z)^{3-j}, \quad 0 \leq i \leq j \leq 3,
\]
so \( S(z) \) can be written as:
\[
S(z) =
\begin{bmatrix}
1 & 0 & 0 & 0 \\
1 - z & z & 0 & 0 \\
(1 - z)^2 & 2(1 - z)z & z^2 & 0 \\
(1 - z)^3 & 3(1 - z)^2z & 3(1 - z)z^2 & z^3 \\
(1 - z)^3 & 3(1 - z)^2z & 3(1 - z)z^2 & z^3 \\
0 & (1 - z)^2 &2(1 - z)z & z^2 \\
0 & 0 & 1-z & z \\
0 & 0 & 0 & 1
\end{bmatrix}
\]

\begin{figure}
    \centering
    \includegraphics[width=1.0\linewidth]{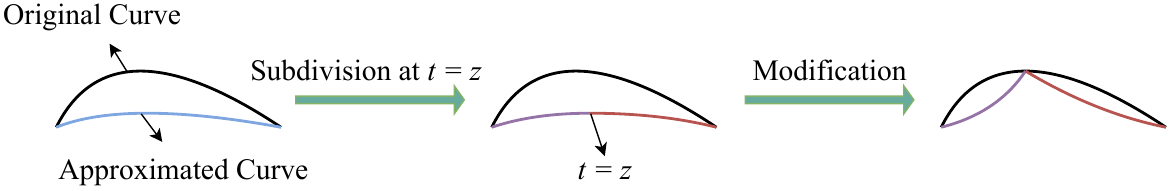}
    \caption{The process of subdivision and modification.
    }
    \label{fig:cubic bezier re-subdivision}
\end{figure}

At the end of each subdivision, the control polygons will be modified. To be more specific, 
we compute the maximum $L_1$-error, along with the corresponding parameter $z$ and point $p_{z}$. The parameter $z$ is then reparameterized to its equivalent parameter on the original Bezier curve, where it is evaluated to obtain the corresponding point $p_{original}$. Finally, $p_{z}$ is updated to $p_{original}$ to ensure the subdivided segments more accurately approximate the original curve, as illustrated in Fig.~\ref{fig:cubic bezier re-subdivision}. 
If the maximum $L_1$-error corresponds to multiple parameters, the segments at all corresponding positions will be subdivided simultaneously. This process of subdivision and modification continues until all segments satisfy the tolerance.

\begin{figure}
    \centering
    \includegraphics[width=1.0\linewidth]{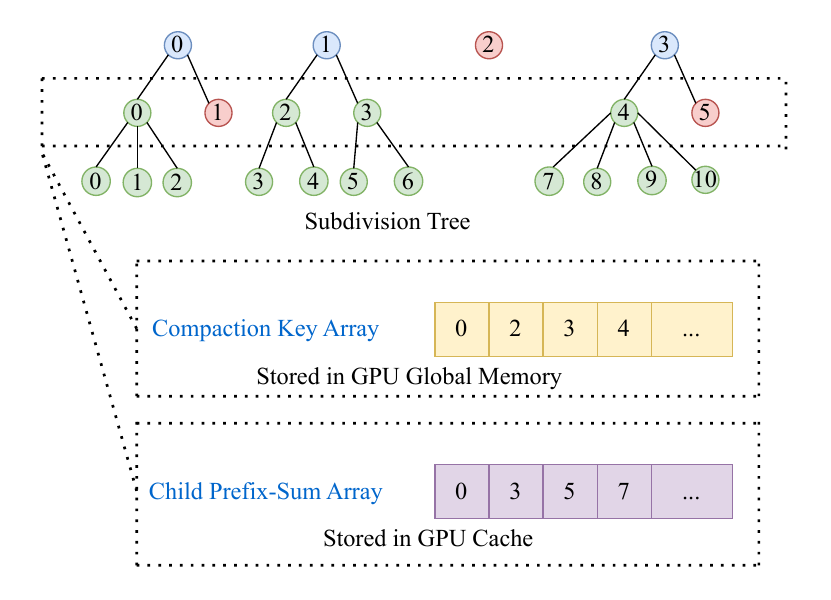}
    \caption{The structure of the subdivision tree.}
    \label{fig:re_subdivision}
\end{figure}

\textbf{GPU optimization.}
In our approximation process, frequently reading the newly subdivided segments's parent information is computationally expensive. To address this, inspired by \cite{yan2019harmonia}, we employ a child prefix sum array and a compaction key array to better align with the GPU memory hierarchy, as shown in Fig.~\ref{fig:re_subdivision}. This approach allows us to utilize a thrust-inclusive scan to further accelerate each level of the subdivision process. In addition, since this step requires multiple kernel restarts, a scheduling algorithm is designed to determine the number of subdivisions to perform each time the kernel starts. The number of subdivisions per execution $K$ is determined by the following formula:
\begin{equation}
K = \min \left\{ \frac{REG_{SM}}{REG_{Sub}} * NUM_{SM}, NUM_{TotalSub} \right\}
\label{subdivision}
\end{equation}
where $REG_{Sub}$ means the registers each subdivision needs, $REG_{SM}$ means the registers each SM has, $NUM_{SM}$ means the total SMs the GPU has, and $NUM_{TotalSub}$ means the total number of segments that need to be subdivided again.
If the current number of segments to be subdivided is too large, the excess segments will be saved for the next execution. This can prevent register spill~\cite{jeon2015gpu} from increased memory divergence due to frequent access to global memory. The approximation process is dynamically scheduled on the GPU until all approximated segments satisfy the specified tolerance.

\textbf{Special case: B-splines of degree 3 and below.}
For B-splines of degree 2 or 3, the operations are further simplified. For the quadratic B-splines, we only need to perform a degree elevation operation after decomposing into piecewise Béziers. For the cubic B-splines, no further processing is required.

\subsubsection{Monotonic decomposition}
\label{monotonic}
Consider the squared distance function from a test point $\mathbf{q}$ to a cubic Bézier curve $C(t)$, which is defined as:
\begin{equation}
D(t) = \|C(t) - \mathbf{q}\|^2
\label{Dt}
\end{equation}
the first derivative of $D(t)$ with respect to $t$ can be expressed as:
\begin{equation}
E(t) = \frac{dD(t)}{dt} = 2(C(t) - \mathbf{q}) \cdot \frac{dC(t)}{dt}
\label{Et}
\end{equation}
similarly, the second derivative of $t$ is
\begin{equation}
E'(t) = \frac{dE(t)}{dt} = 2 \left[ \left(\frac{dC(t)}{dt}\right)^2 + (C(t) - \mathbf{q}) \cdot \frac{d^2C(t)}{dt^2} \right]
\label{E't}
\end{equation}

Based on Sec.~\ref{error-controlled decomposition}, a method is proposed for point projection and inversion, where approximated cubic Bézier curves are decomposed into monotonic segments, facilitating the identification of global optima.
Here `monotonic' means that $E(t)$ is monotonic within the interval [0,1], 
ensuring that each segment contains at most one potential projection point.

Given that $C(t)$ is a cubic Bézier curve, its second derivative $E'(t)$ results in a polynomial of degree 4. Using Ferrari's method~\cite{strang2022introduction}, we can analytically compute the roots of $E'(t)$ within the interval $[0,1]$. If any roots are found, they serve as subdivision parameters for the cubic Bézier curve. Once the curve has been subdivided at all locations where $E'(t) = 0$, the remaining segments are guaranteed to be monotonic.

Because the analytical method avoids generating local solutions and ensures that all threads within a warp execute the same number of operations, the computation of the roots of $E'(t)$ on GPU is significantly faster and more stable compared to traditional numerical methods. Experiments show that the analytical method is about 4 times faster than Newton's iteration.

\begin{figure}
    \centering
    \includegraphics[width=1.0\linewidth]{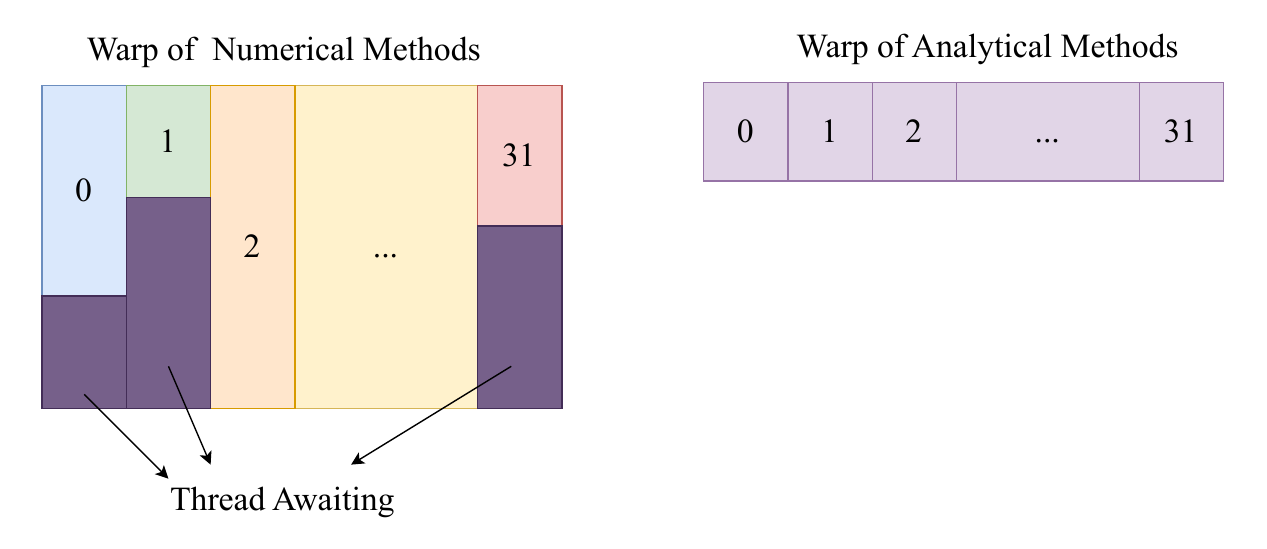}
    \caption{Execution status of threads within a warp for numerical (left) and analytical methods (right).}
    \label{fig:compare_ana_newton}
\end{figure}

\subsection{M-rep for Bézier projection/inversion}
\label{M-rep for bezier inversion/projection}
With the help of monotonic segments, robust projection and inversion operations can be performed. However, projection and inversion themselves often involve complex iterative calculations, making them unsuitable for matrix operations on GPUs. To bridge the gap, 
$E(t)$ is transformed into a non-parametric Bézier form after applying an elimination criterion, then Bézier clipping is used to locate its intersection with the x-axis. This process converts the problem of solving polynomial equations in projection and inversion into a line segment-line intersection problem, which is more suitable for GPU parallel computation. Additionally, we show that the process of converting $E(t)$ from polynomial form to non-parametric Bézier form can be converted into matrix operations, and the data structure can be properly designed to utilize tensor core acceleration.

\begin{figure}
    \centering
    \includegraphics[width=1.0\linewidth]{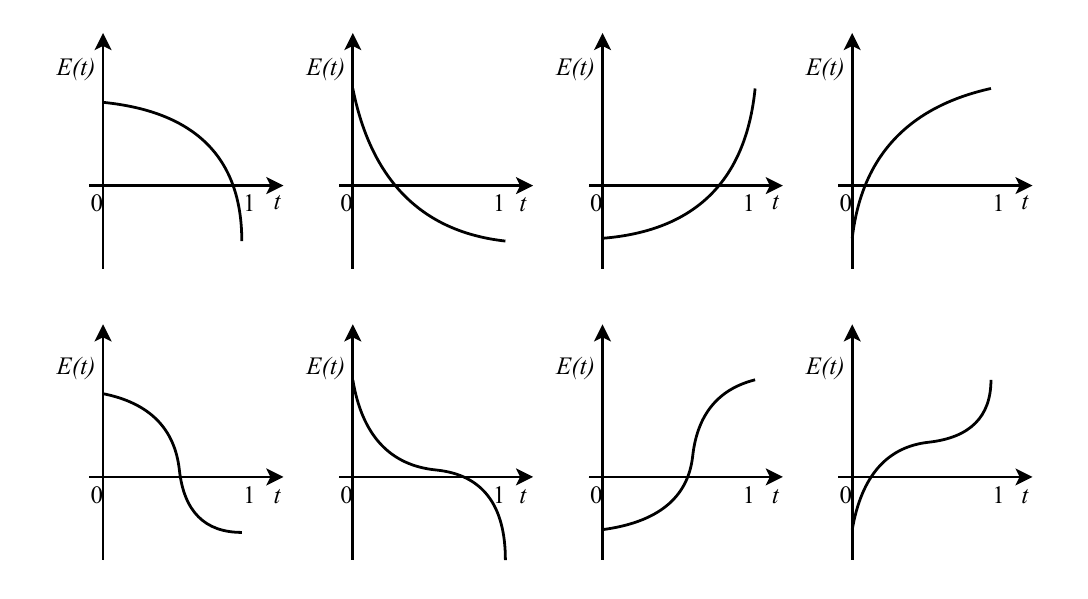}
    \caption{Possible cases of $E(t)$.}
    \label{fig:Et}
\end{figure}

\textbf{Elimination criterion.}
Since there are actually no potential projection points on most monotonic segments, it is necessary to design an elimination criterion to reduce the subsequent computational cost. As in Fig.~\ref{fig:Et}, from the monotonicity of E(t), we obtain the following properties:

\textit{Property 1.} If $E(0) >= 0$ or $E(0) * E(1) > 0$, the corresponding monotonic segment will not have potential projection points, unless it contains the first or last endpoint of the original B-spline curve.

\textit{Property 2.} If the monotonic segment contains the first or last endpoint of the original B-spline curve and satisfies \textit{property 1}, the projection point can only exist on the endpoint contained in this monotonic segment.

\textit{Proof.} When $E(0)$ is greater than 0, there can only be extremely large values of $D(t)$, not extremely small ones. And when $E(0) * E(1) > 0$, there are no roots of $E(t)$ on the interval $[0,1]$, and the corresponding $D(t)$ has no extrema. In both cases, the projection points can only appear at the first or last endpoint of the original B-spline curve.

\begin{figure}
    \centering
    \includegraphics[width=0.95\linewidth]{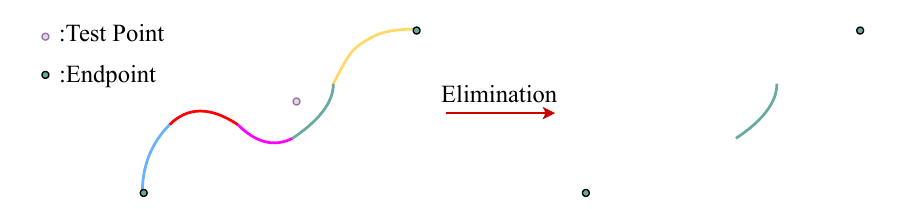}
    \caption{Illustration of the elimination criterion. Only the segments containing the local minimum of $D(t)$ and the endpoints are retained.}
    \label{fig:elinimation_criteria}
\end{figure}

Based on these two properties, the majority of monotonic segments can be eliminated. The remaining monotonic segments along with the endpoints are then sent to the next kernel for further processing. Our process can be seen in Fig.~\ref{fig:elinimation_criteria}.

\begin{figure}
    \centering
    \includegraphics[width=1.0\linewidth]{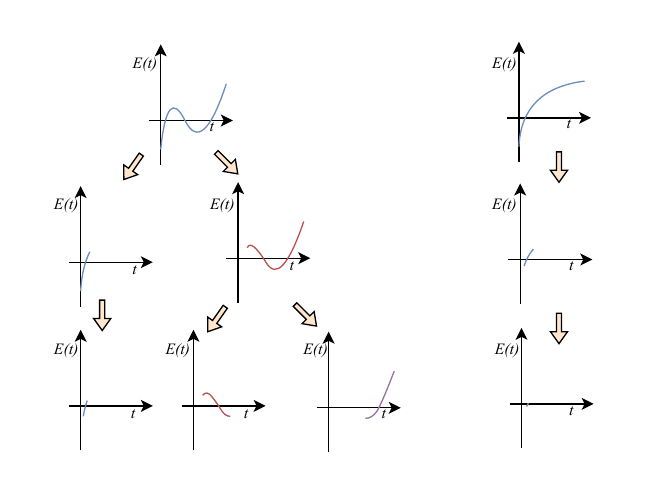}
    \caption{The GPU execution process of Bézier clipping, where $E(t)$ is non-monotonic (left) and monotonic (right).}
    \label{fig:monotonic}
\end{figure}

\textbf{Rebasement.}
After eliminating most of the monotonic segments that do not contain solutions, we convert $E(t)$ into non-parametric Bézier representations and then use Bézier clipping to find its intersection with the x-axis. This choice is motivated by the fact that in Bézier clipping, 
if $E(t)$ has multiple intersections with the x-axis, the curve needs to be subdivided in half and to compute the intersections of each half with the x-axis~\cite{sederberg1990curve}.
Each subdivision in Bézier clipping will generate two new Bézier clipping tasks, while $E(t)$ is monotonic can ensure that there will be no need to subdivide the curve, which can greatly reduce warp divergence, as illustrated in Fig.~\ref{fig:monotonic}.

\textbf{Tensor core acceleration.}
The relationship between power basis and Bernstein basis for an \(n\)-degree polynomial $E(t)$ is given by:
\[
\mathbf{b} = T \mathbf{a},
\]
where $\mathbf{a} = [a_0, a_1, \dots, a_n]^T,$ is the power basis coefficient vector of $E(t)$. And $\mathbf{b} = [b_0, b_1, \dots, b_n]^T,$ is the Bernstein basis coefficient vector of $E(t)$. And \(T\) is the transformation matrix, elements in $T$ can be computed as:
\[
T_{i,j} =
\begin{cases}
\binom{i}{j} \cdot \binom{n}{i}^{-1}, & \text{if } j \leq i, \\
0, & \text{otherwise}.
\end{cases}
\]

In our cases, $E(t)$ is of $5$-degree, in this scenario:
\[
T =
\begin{bmatrix}
1 & 0 & 0 & 0 & 0 & 0 \\
1 & 0.2 & 0 & 0 & 0 & 0 \\
1 & 0.4 & 0.1 & 0 & 0 & 0 \\
1 & 0.6 & 0.3 & 0.1 & 0 & 0 \\
1 & 0.8 & 0.6 & 0.4 & 0.2 & 0 \\
1 & 1 & 1 & 1 & 1 & 1
\end{bmatrix}.
\]

Since T is always a constant, we can merge all $\mathbf{a}$ into a large matrix, then use tensor core to efficiently accelerate the computation and obtain all $\mathbf{b}$ at once. For more details, please see the supplementary materials.

\subsection{GPU-optimized projection/inversion}
\label{WCWS(warp cooperative work sharing) bezier clipping}
To fully utilize GPU resources during the projection and inversion process, especially in large-scale tasks, we first introduce a warp-centric Bézier clipping method to evenly distribute the GPU workload. Then, a point collaborative work sharing (PCWS) strategy is proposed, which uses a partitioning-based grouping approach and atomic locks to accurately and efficiently update the final projection points.

\subsubsection{Warp-centric projection/inversion}
\label{sec:warp-centric projection/inversion}
\begin{figure}
    \centering
    \includegraphics[width=1.0\linewidth]{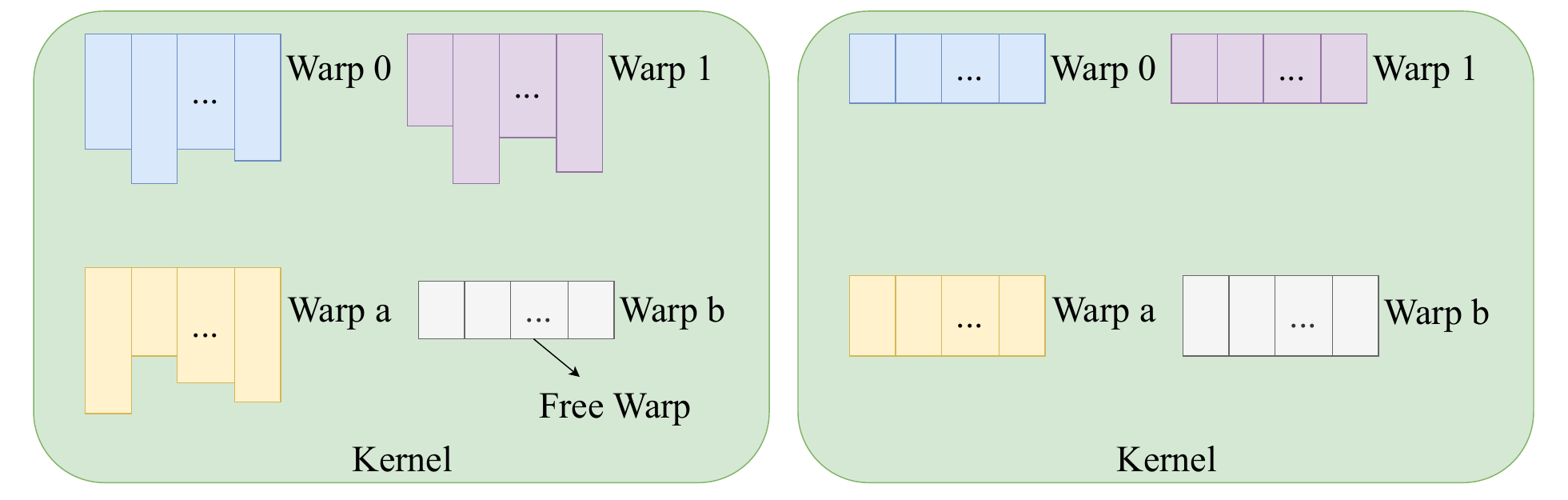}
    \caption{Thread-centric schedule of bezier clipping (left) and warp-centric schedule of bezier clipping (right).}
    \label{fig:warp centric intersection}
\end{figure}

In our Bézier clipping process, a critical step involves calculating the intersections of the convex hull with the x-axis. Instead of adopting a thread-centric approach, where each thread handles the intersection of the convex hull and the x-axis individually, a more efficient warp-centric approach is employed.
To be more specific, we designed an algorithm that dynamically adjusts the number of intersections calculated by each thread. Our formula for the number of intersections $K$ each thread computes can be written as:
\begin{equation}
K = \left\lceil \frac{\sum_{i=0}^{N} NUM_{i} \times ConvexHull_{edge}}{\text{active warp} \times \text{warp size}} \right\rceil
\label{clip}
\end{equation}
where $N$ is the number of all test points, $NUM_{i}$ means the number of segments with potential projection points per test point $\mathbf{q}_{i}$ has, $ConvexHull_{edge}$ means the number of edges per convex hull has, which is 6 in our case. Active warp is the maximum number of warps that the GPU supports for parallel computing, and warp size is the number of threads in a warp, which is a fixed number (32 in Nvidia).
This scheduling algorithm can avoid imbalanced workloads, where some warps are overloaded while others remain idle, as shown in Fig.~\ref{fig:warp centric intersection}.

After obtaining the intersections of the convex hull and the x-axis, we will clip the curve. According to Eq.~\eqref{subLandR}, our clipping process can also be expressed in matrix representation as follows:
\begin{equation}
S_{[z_1, z_2]} = S_{L5}\left(\frac{z_2 - z_1}{1 - z_1}\right) \cdot S_{R5}(z_1)
\label{subdivisionmatrix}
\end{equation}
where $z_1$ and $z_2$ are the left and right intersections of the convex hull and the x-axis on [0,1], $S_{L5}$ and $S_{R5}$ are the corresponding subdivision matrix when degree is $5$. And $S_{[z_1, z_2]}$ means the clipping matrix that clips the curve from $z_1$ to $z_2$. The experimental results indicate that the parameters converge to approximately $1\mathrm{e}{-6}$ after performing Bézier clipping three times.

\subsubsection{Point cooperative work sharing strategy}
\label{PCWS}
After performing Bézier clipping, multiple groups of potential projection points are obtained. For each group, the task is to identify the point with the shortest distance as the final projection point. This large-scale, small-batch task presents significant challenges in parallelization due to memory contention when multiple threads attempt to simultaneously update the minimum value. A naive approach that iterates over each group of potential projection points for a given $\mathbf{q}_{i}$, leads to low parallel efficiency. Alternatively, grouping the potential projection points for each $\mathbf{q}_{i}$ into a warp for warp-level reduction via warp shuffle is not optimal, as the number of potential points in each group often does not align with the warp size, causing thread inefficiencies within the warp.

\begin{figure}
    \centering
    \includegraphics[width=1.0\linewidth]{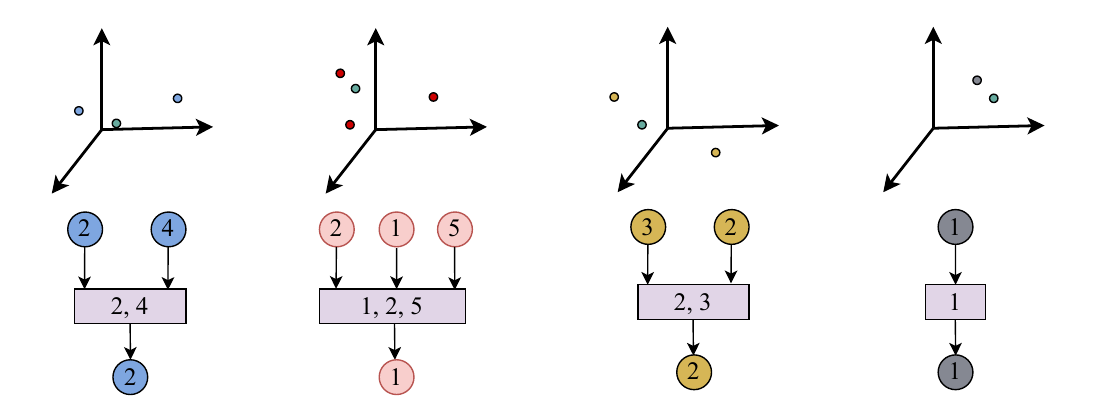}
    \caption{Illustration of PCWS minimum distance detection.}
    \label{fig:PCWS}
\end{figure}

To address these issues, inspired by WCWS \cite{awad2019engineering}, we propose a point cooperative work sharing strategy. However, unlike WCWS, which processes irregular data at the warp level, our method focuses on handling each group of potential projection points as a unit. To be more specific, the endpoints selected by the elimination criterion, along with the points identified through Bézier clipping, are considered as potential projection points. The distances between these potential projection points and their corresponding $\mathbf{q}_{i}$ are then computed in parallel. After that, atomic locks are employed to safely update the shortest distance for each group. This method ensures the correctness of the final projection points while maintaining efficient GPU parallelism by minimizing memory contention and reducing overhead, as illustrated in Fig.~\ref{fig:PCWS}.

Regarding the inversion, since our curves retain the information of the original B-spline curve during the approximation process, the corresponding parameters can be obtained through reparameterization.

\section{Experiments}
\label{sec: results}
We implemented our method and performed comparisons with open-source software SISL~\cite{sisl} and commercial 3D modeler Parasolid~\cite{parasolid}.
The open-source code from SISL is directly utilized, while the Parasolid experiment was developed through secondary development on NX software.
All experiments were executed on an Nvidia RTX 3090 GPU paired with an Intel Core i9-12900K CPU running at 5.20GHz server. The GPU is equipped with 82 SMs, 128 KB of L1 cache, 6 MB of L2 cache, and a total of 24 GB of global memory. The method has been implemented using C++ 17 and CUDA 12.2, operating within the Ubuntu 20.04 environment and Windows 10 in a virtual machine environment. $L_1$-error tolerance $\alpha$ of M-rep and geometry resolution of SISL are set to $1\mathrm{e}{-4}$.

\subsection{Examples}
\def\tabularxcolumn#1{m{#1}}
\begin{table*}[htb]
\caption{Statistics of each module using the M-rep method with $10^{4}$ projected points in 2D cases.}
\centering
\renewcommand{\arraystretch}{1.0}  
\setlength{\extrarowheight}{0pt}  
\setlength{\tabcolsep}{4pt}  
\begin{tabularx}{1.0\textwidth}{
    >{\arraybackslash}>{\hsize=.8\hsize\linewidth=\hsize}X   
    >{\centering\arraybackslash}>{\hsize=.25\hsize\linewidth=\hsize}X  
    >{\centering\arraybackslash}>{\hsize=.25\hsize\linewidth=\hsize}X
    >{\centering\arraybackslash}>{\hsize=.25\hsize\linewidth=\hsize}X
    >{\centering\arraybackslash}>{\hsize=.25\hsize\linewidth=\hsize}X
    >{\centering\arraybackslash}>{\hsize=.25\hsize\linewidth=\hsize}X
    >{\centering\arraybackslash}>{\hsize=.25\hsize\linewidth=\hsize}X
    >{\centering\arraybackslash}>{\hsize=.25\hsize\linewidth=\hsize}X
    >{\centering\arraybackslash}>{\hsize=.25\hsize\linewidth=\hsize}X
    >{\centering\arraybackslash}>{\hsize=.25\hsize\linewidth=\hsize}X
   }
    \Xhline{1pt}
    Model &(a) &(b) &(c) &(d) &(e) &(f) &(g) &(h) &(i) \\
    \Xhline{1pt}
    Degree                   &4   &5    &6     &4   &5     &6   &4      &5    &6\\
    Knots Length             &10  &12   &14    &14  &18    &22  &35     &22   &57\\
    Number of Control Points &5   &6    &7     &9   &12    &15  &30     &16   &50\\
    Decomposition Time (ms) &1.52 &9.00 &10.9 &4.73 &2.11 &7.21 &3.65 &2.59 &8.65\\
    Projection Time (ms)      &0.57 &0.51 &1.30 &0.93 &0.61 &0.64 &1.68 &0.78 &1.20\\
    Total Time (ms)           &2.09 &9.51 &12.2 &5.67 &2.72 &7.85 &5.32 &3.37 &9.85\\
    Average Time ($\mu$s)     &0.21 &0.95 &1.22 &0.57 &0.27 &0.78 &0.53 &0.34 &0.98\\
    \Xhline{1pt}
\end{tabularx}
\label{tab: 2D_case}
\end{table*}

\def\tabularxcolumn#1{m{#1}}
\begin{table*}[htb]
\caption{Statistics of each module using the M-rep method with $10^{4}$ projected points in 3D cases.}
\centering
\renewcommand{\arraystretch}{1.0}  
\setlength{\extrarowheight}{0pt}  
\setlength{\tabcolsep}{4pt}  
\begin{tabularx}{1.0\textwidth}{
    >{\arraybackslash}>{\hsize=.8\hsize\linewidth=\hsize}X   
    >{\centering\arraybackslash}>{\hsize=.25\hsize\linewidth=\hsize}X  
    >{\centering\arraybackslash}>{\hsize=.25\hsize\linewidth=\hsize}X
    >{\centering\arraybackslash}>{\hsize=.25\hsize\linewidth=\hsize}X
    >{\centering\arraybackslash}>{\hsize=.25\hsize\linewidth=\hsize}X
    >{\centering\arraybackslash}>{\hsize=.25\hsize\linewidth=\hsize}X
    >{\centering\arraybackslash}>{\hsize=.25\hsize\linewidth=\hsize}X
    >{\centering\arraybackslash}>{\hsize=.25\hsize\linewidth=\hsize}X
    >{\centering\arraybackslash}>{\hsize=.25\hsize\linewidth=\hsize}X
    >{\centering\arraybackslash}>{\hsize=.25\hsize\linewidth=\hsize}X
   }
    \Xhline{1pt}
    Model &(j) &(k) &(l) &(m) &(n) &(o) &(p) &(q) &(r) \\
    \Xhline{1pt}
    Degree                   &4   &5    &6     &4   &5     &6   &4      &5    &6\\
    Knots Length             &10  &12   &14    &14  &18    &22  &35     &46   &57\\
    Number of Control Points &5   &6    &7     &9   &12    &15  &30     &40   &50\\
    Decomposition Time (ms) &2.82 &5.93 &12.4 &5.71 &1.98 &9.91 &3.62 &3.60 &10.9\\
    Projection Time (ms)      &0.61 &0.58 &1.00 &0.60 &0.61 &0.99 &1.01 &1.44 &1.15\\
    Total Time (ms)           &3.43 &6.52 &13.4 &6.30 &2.61 &10.9 &4.63 &5.04 &12.1\\
    Average Time ($\mu$s)     &0.34 &0.65 &1.34 &0.63 &0.26 &1.09 &0.46 &0.50 &1.21\\
    \Xhline{1pt}
\end{tabularx}
\label{tab: 3D_case}
\end{table*}

\begin{figure}[htb]
    \centering
    \subfigure[]{
        \includegraphics[width=0.13\textwidth]{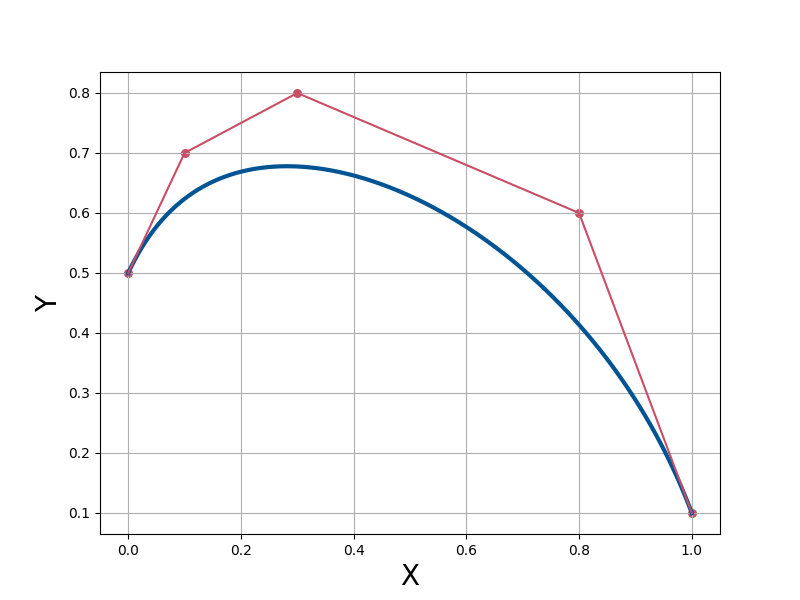}  
        \label{fig:case1}
    }
    \subfigure[]{
        \includegraphics[width=0.13\textwidth]{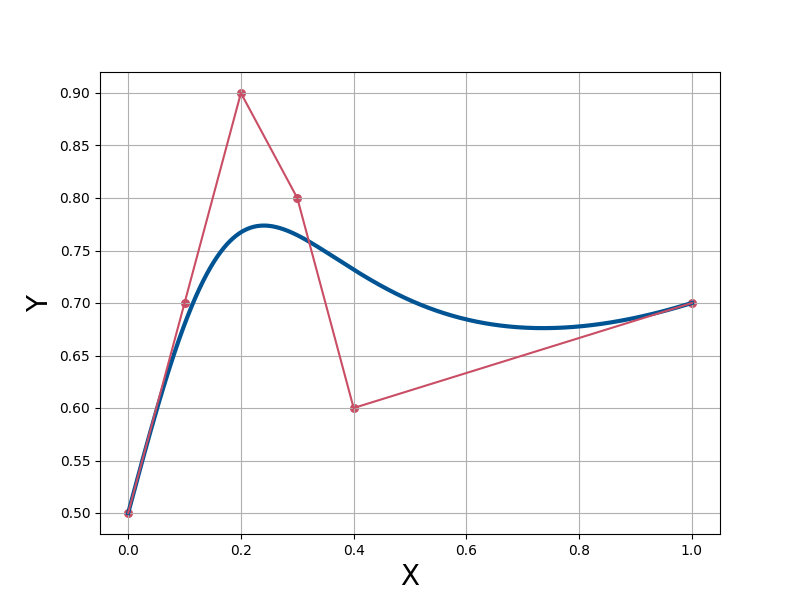}  
        \label{fig:case2}
    }
    \subfigure[]{
        \includegraphics[width=0.13\textwidth]{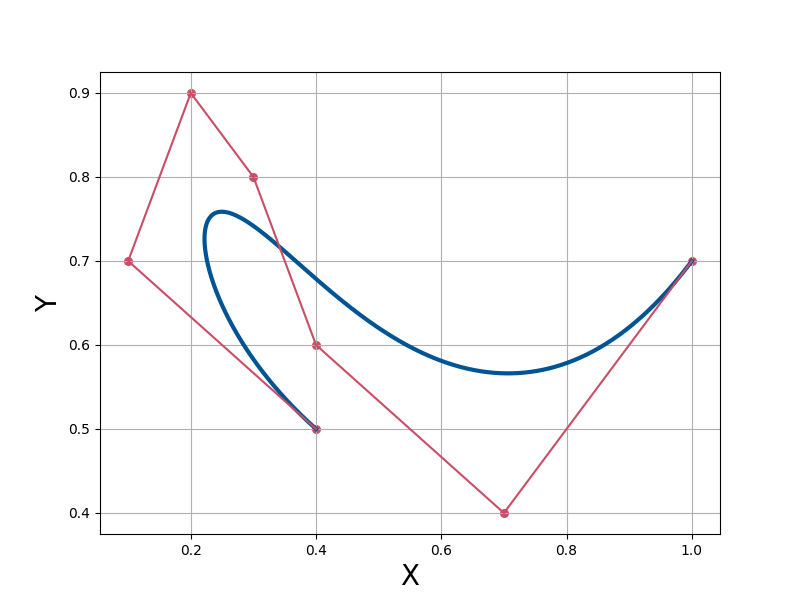}  
        \label{fig:case3}
    }

    \subfigure[]{
        \includegraphics[width=0.13\textwidth]{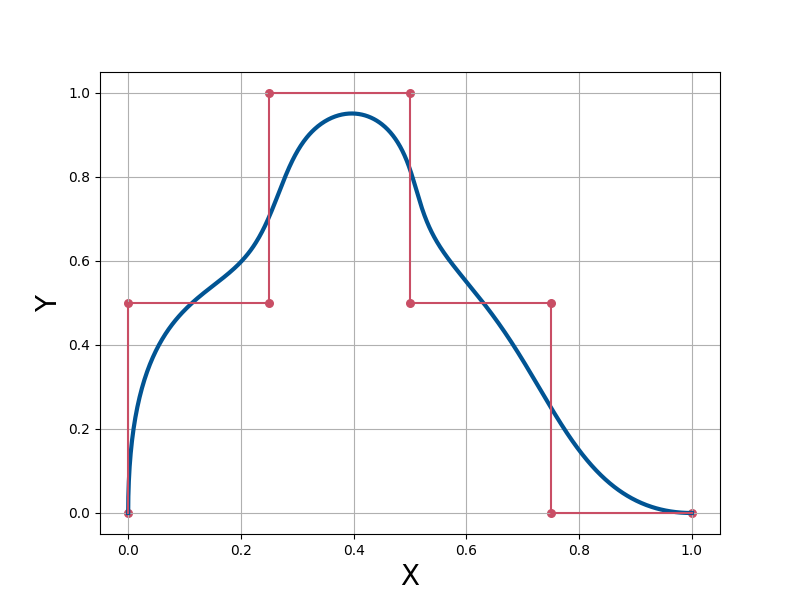}  
        \label{fig:case4}
    }
    \subfigure[]{
        \includegraphics[width=0.13\textwidth]{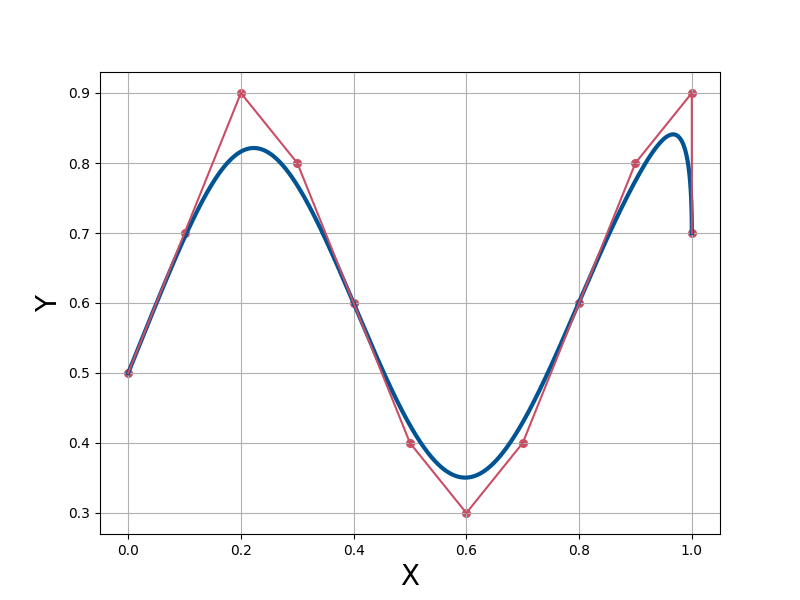}  
        \label{fig:case5}
    }
    \subfigure[]{
        \includegraphics[width=0.13\textwidth]{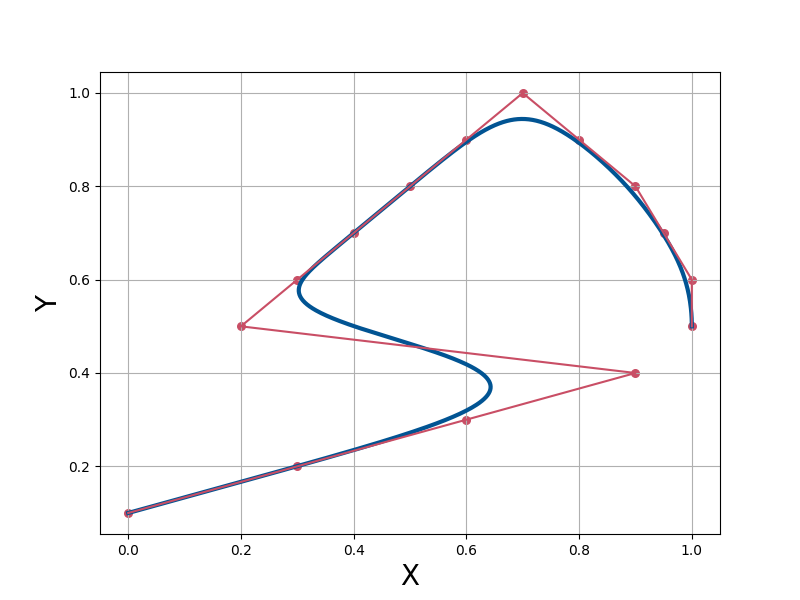}  
        \label{fig:case6}
    }

    \subfigure[]{
        \includegraphics[width=0.13\textwidth]{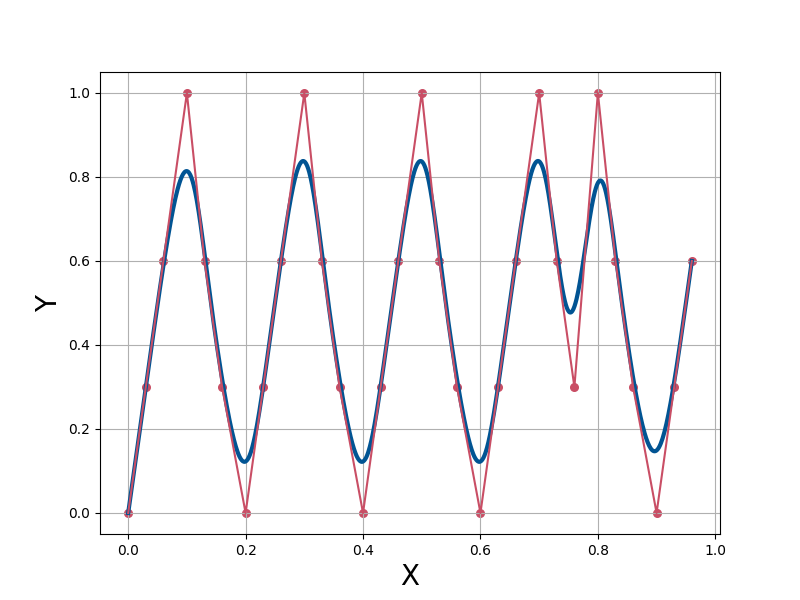}  
        \label{fig:case7}
    }
    \subfigure[]{
        \includegraphics[width=0.13\textwidth]{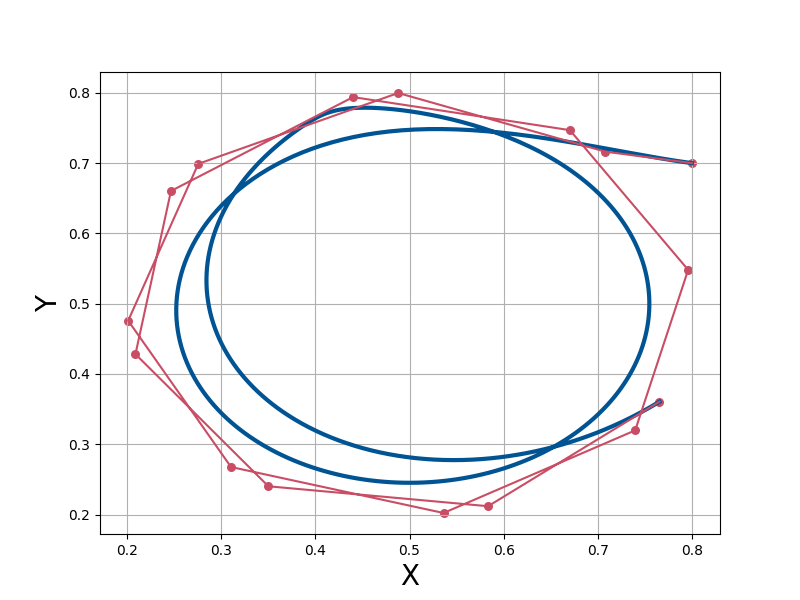}  
        \label{fig:case8}
    }
    \subfigure[]{
        \includegraphics[width=0.13\textwidth]{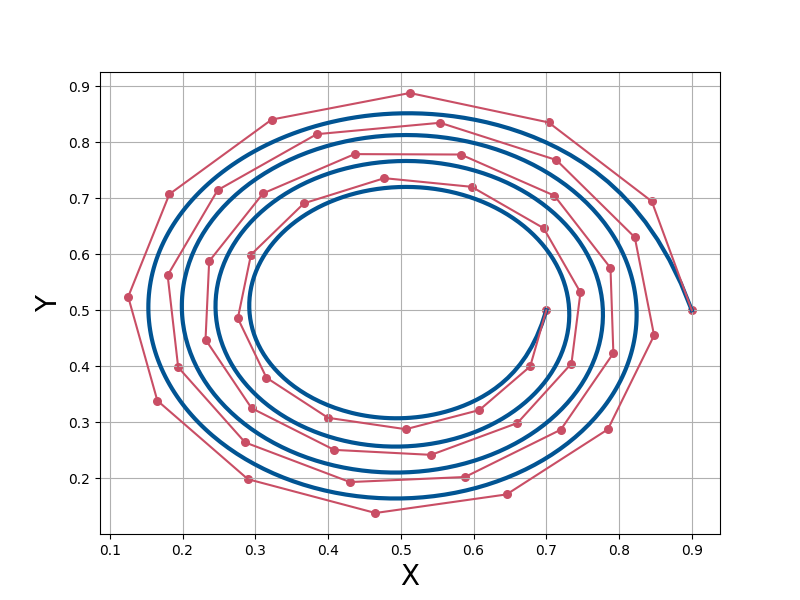}  
        \label{fig:case9}
    }

    \caption{2D cases for projection and inversion experiments.}
    \label{fig:2D_case}
\end{figure}

\begin{figure}[htb]
    \centering
    \setcounter{subfigure}{9}
    \subfigure[]{
        \includegraphics[width=0.12\textwidth]{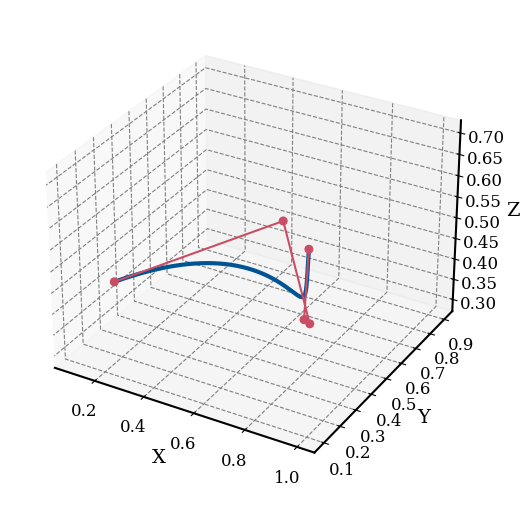}  
        \label{fig:case10}
    }
    \subfigure[]{
        \includegraphics[width=0.12\textwidth]{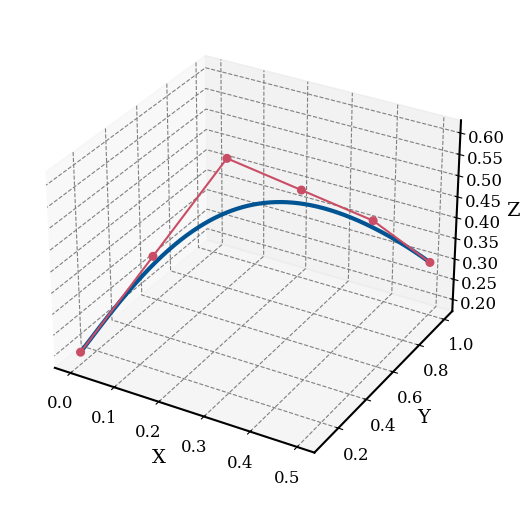}  
        \label{fig:case11}
    }
    \subfigure[]{
        \includegraphics[width=0.12\textwidth]{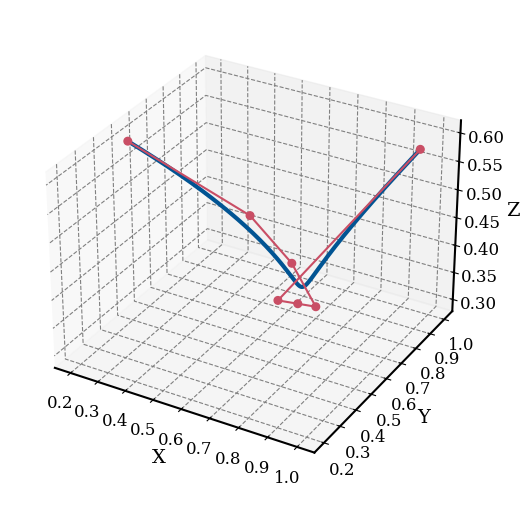}  
        \label{fig:case12}
    }

    \subfigure[]{
        \includegraphics[width=0.12\textwidth]{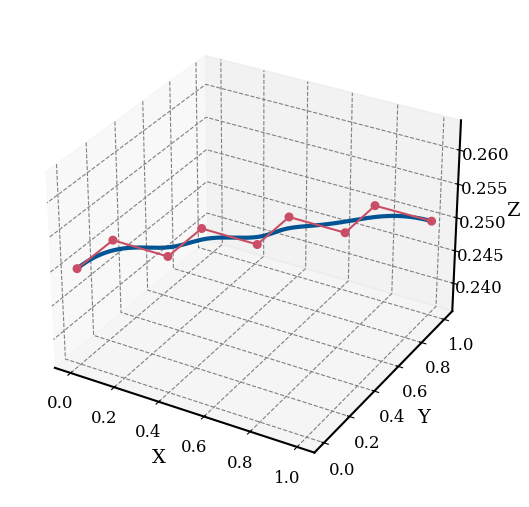}  
        \label{fig:case13}
    }
    \subfigure[]{
        \includegraphics[width=0.12\textwidth]{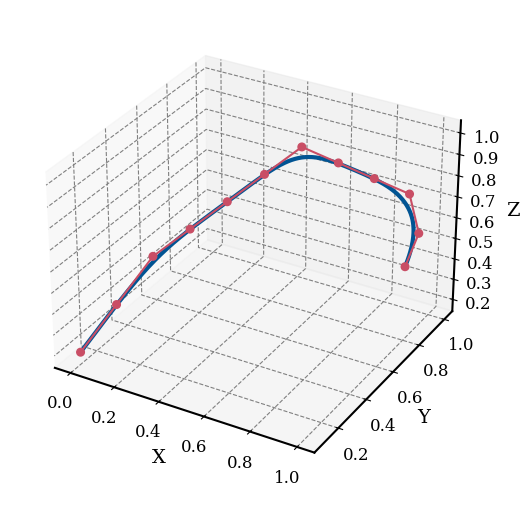}  
        \label{fig:case14}
    }
    \subfigure[]{
        \includegraphics[width=0.12\textwidth]{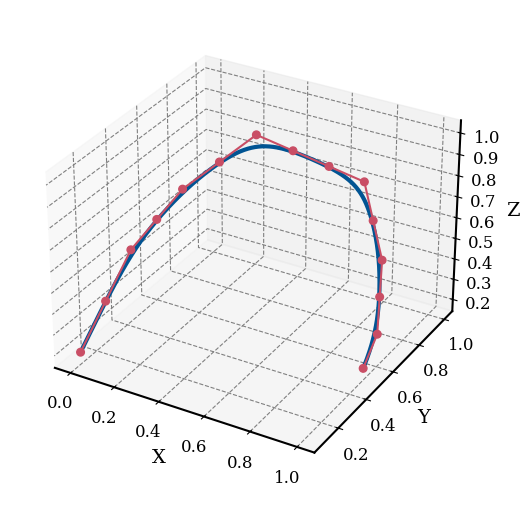}  
        \label{fig:case15}
    }

    \subfigure[]{
        \includegraphics[width=0.12\textwidth]{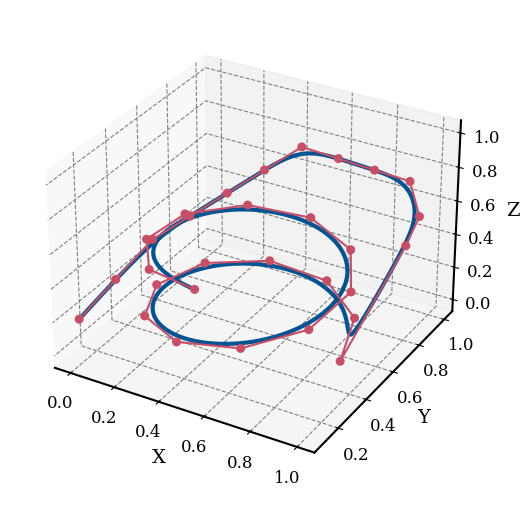}  
        \label{fig:case16}
    }
    \subfigure[]{
        \includegraphics[width=0.12\textwidth]{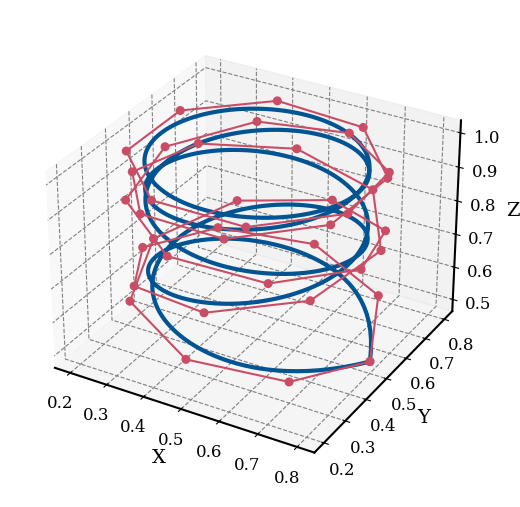}  
        \label{fig:case17}
    }
    \subfigure[]{
        \includegraphics[width=0.12\textwidth]{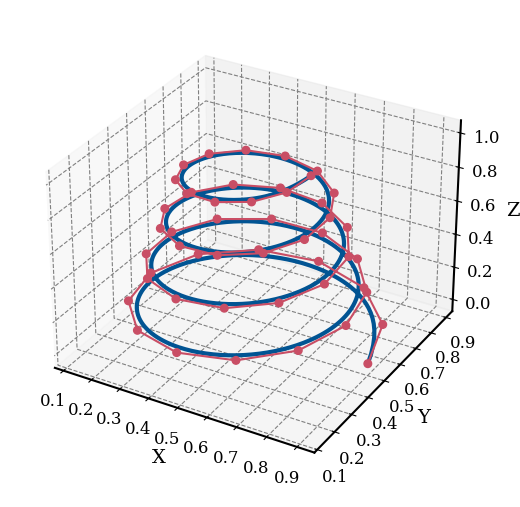}  
        \label{fig:case18}
    }

    \caption{3D cases for projection and inversion experiments.}
    \label{fig:3D_case}
\end{figure}

\begin{figure}
    \centering
    \subfigure[Time distribution of M-rep on Fig.~\ref{fig:case14}.]{
        \includegraphics[width=0.22\textwidth]{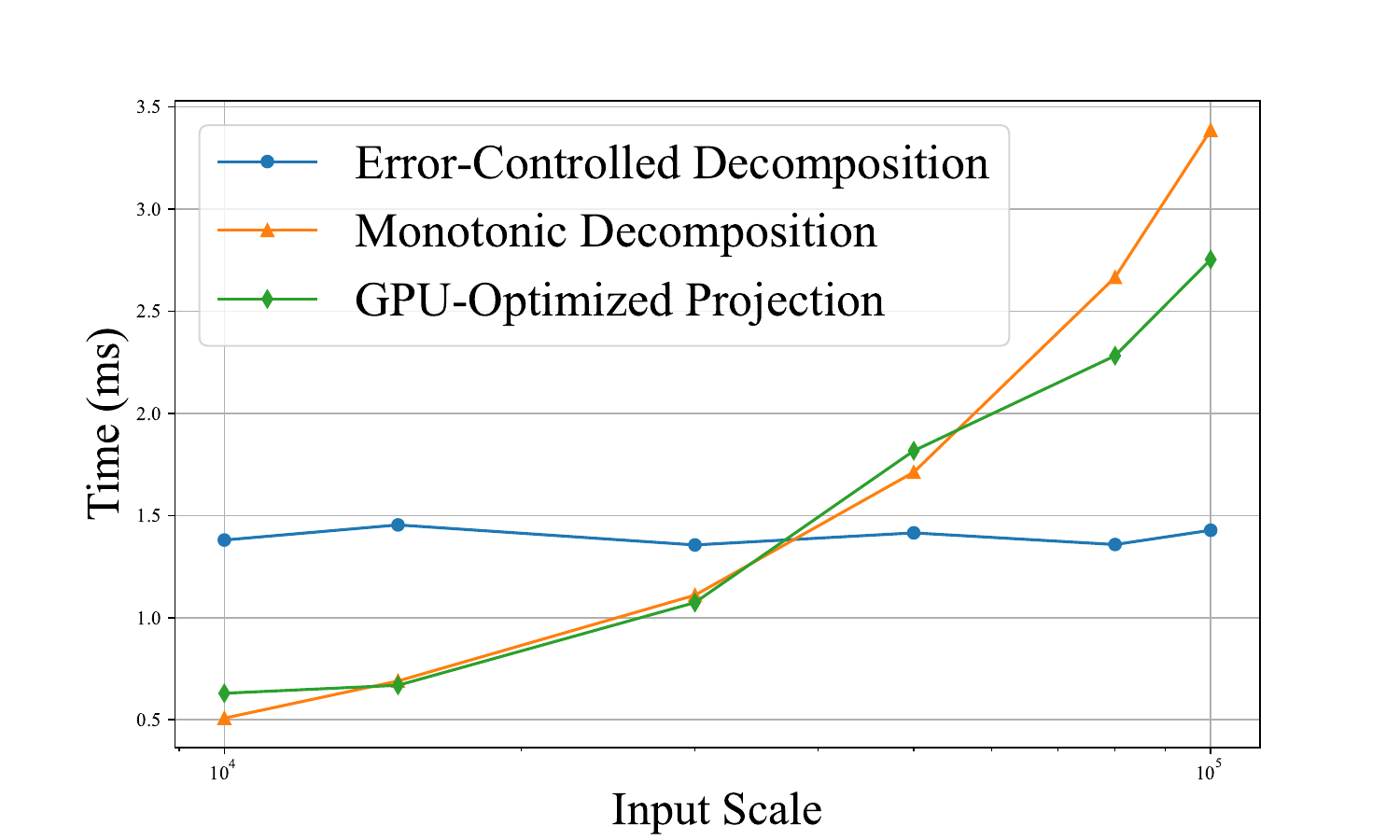} 
        \label{fig:time_distribution_5normal}
    }
    \subfigure[Time distribution of M-rep on Fig.~\ref{fig:case18}.]{
        \includegraphics[width=0.22\textwidth]{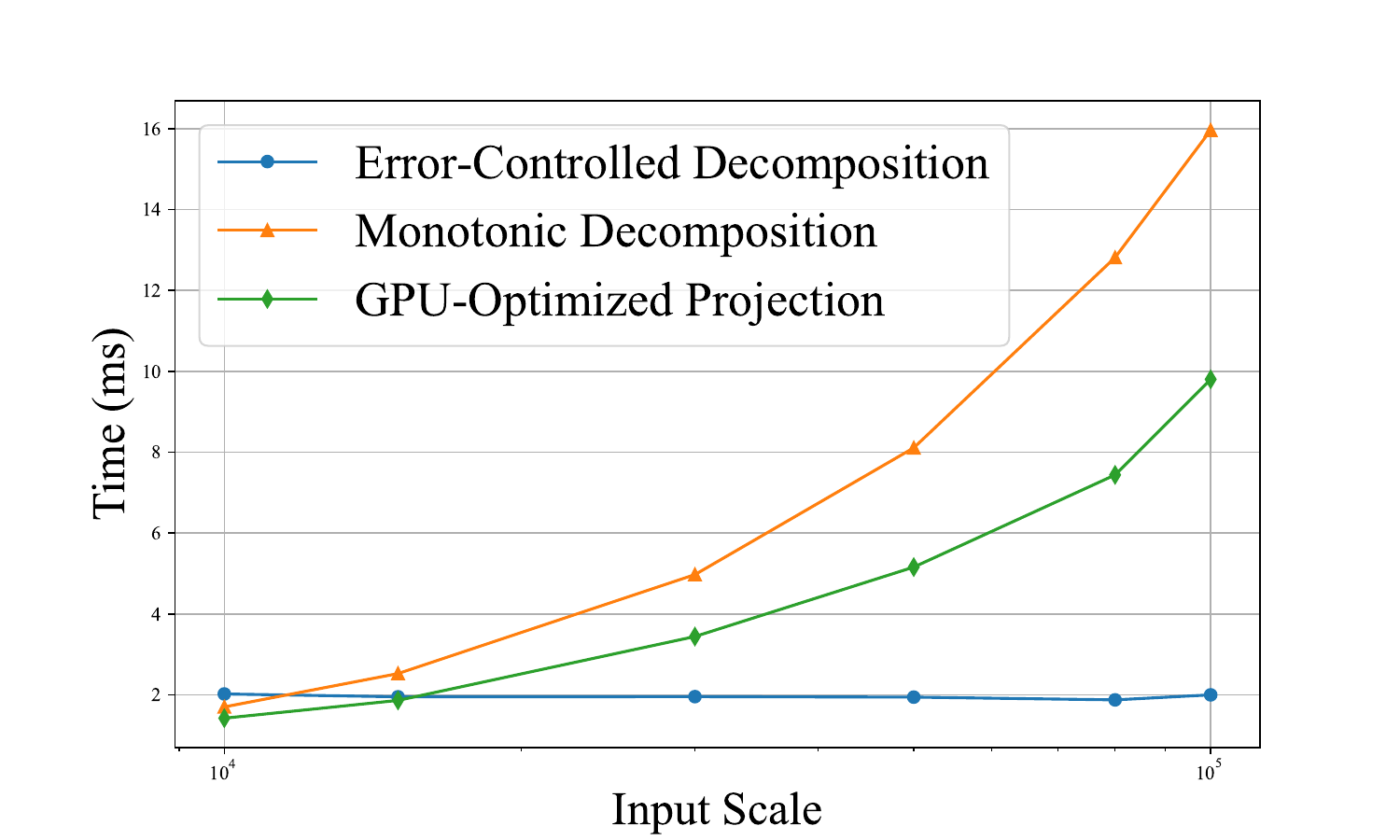}  
        \label{fig:time_distribution_5hard}
    }
    \caption{Time distribution of M-rep on two models with different complexity.}
    \label{fig:time_distribution}
\end{figure}

\begin{figure}
    \centering
    \subfigure[Time comparison on Fig.~\ref{fig:case1}.]{
        \includegraphics[width=0.22\textwidth]{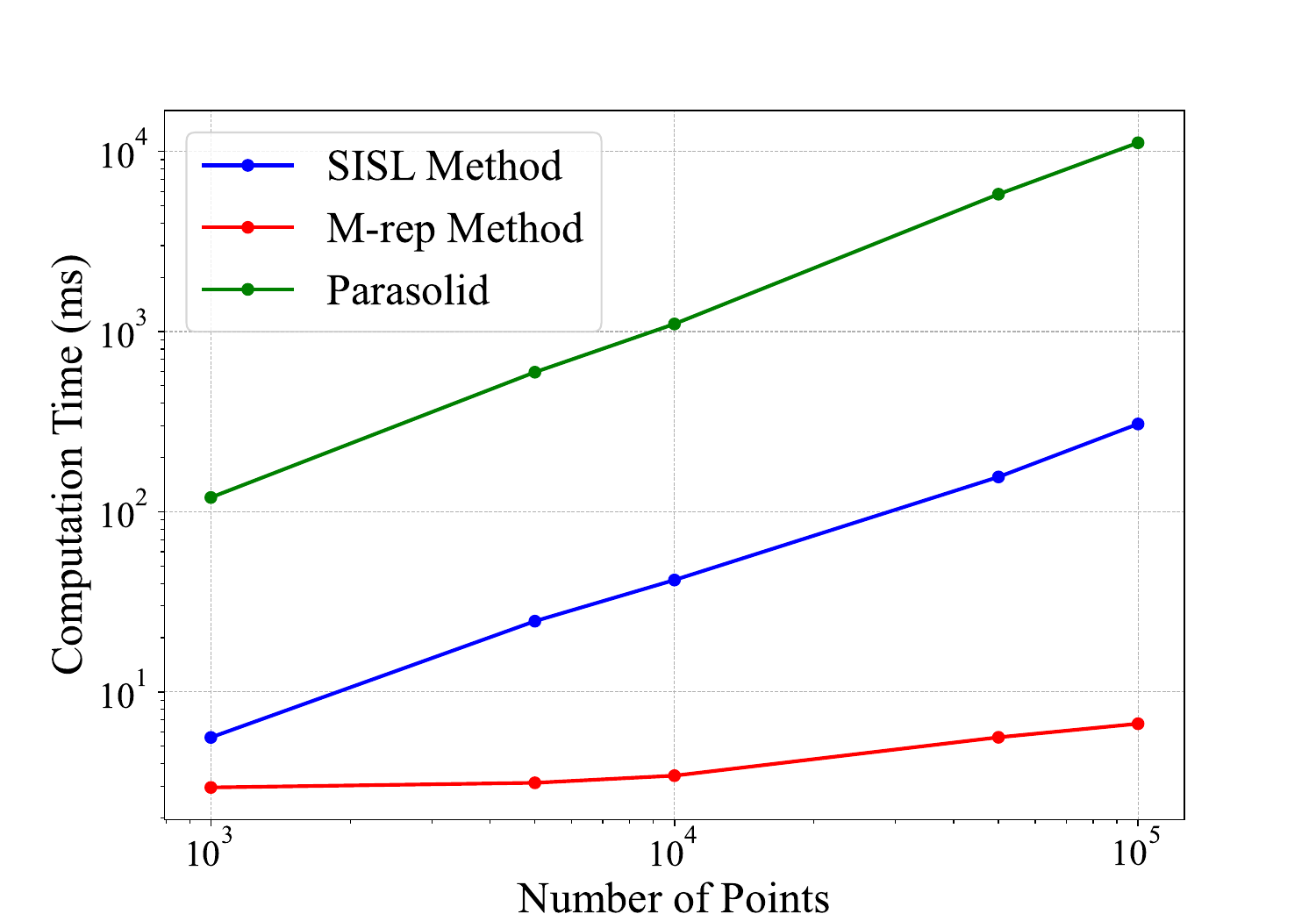} 
        \label{fig:time_comparesion_4easy}
    }
    \subfigure[Time comparison on Fig.~\ref{fig:case14}.]{
        \includegraphics[width=0.22\textwidth]{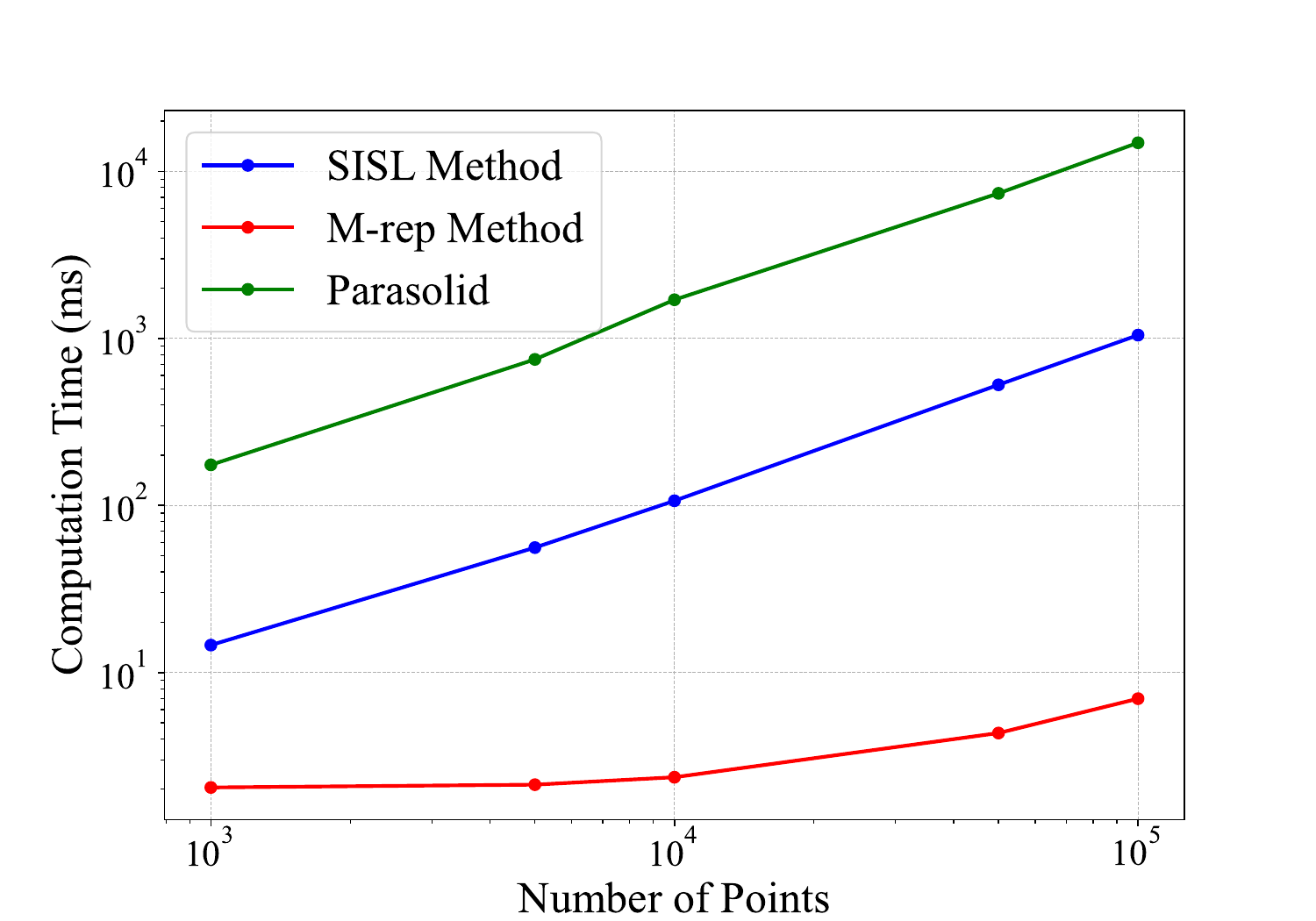}
        \label{fig:time_comparesion_5normal}
    }
    \subfigure[Time comparison on Fig.~\ref{fig:case17}.]{
        \includegraphics[width=0.22\textwidth]{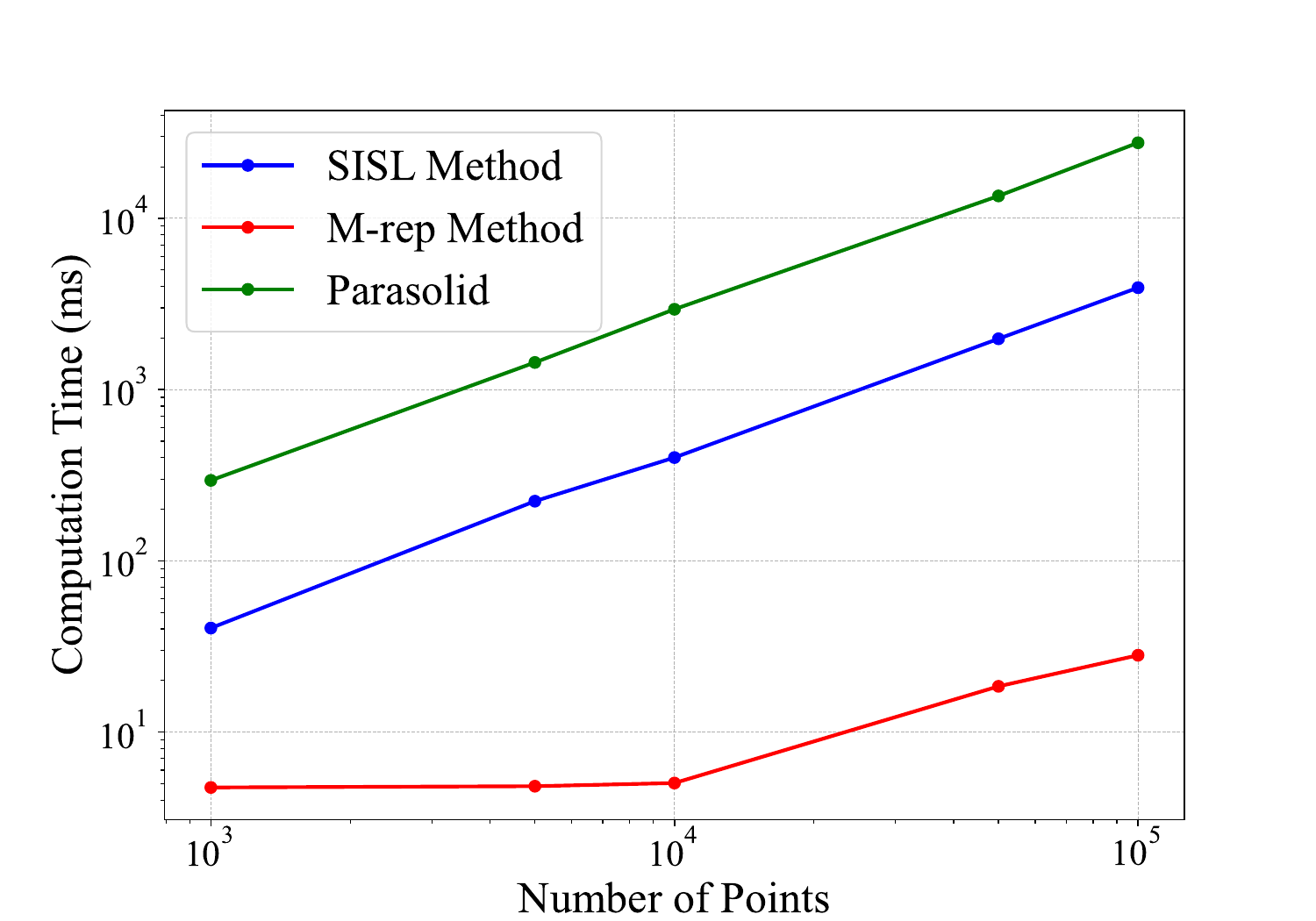}  
        \label{fig:time_comparesion_5hard}
    }
        \subfigure[Time comparison on Fig.~\ref{fig:case18}.]{
        \includegraphics[width=0.22\textwidth]{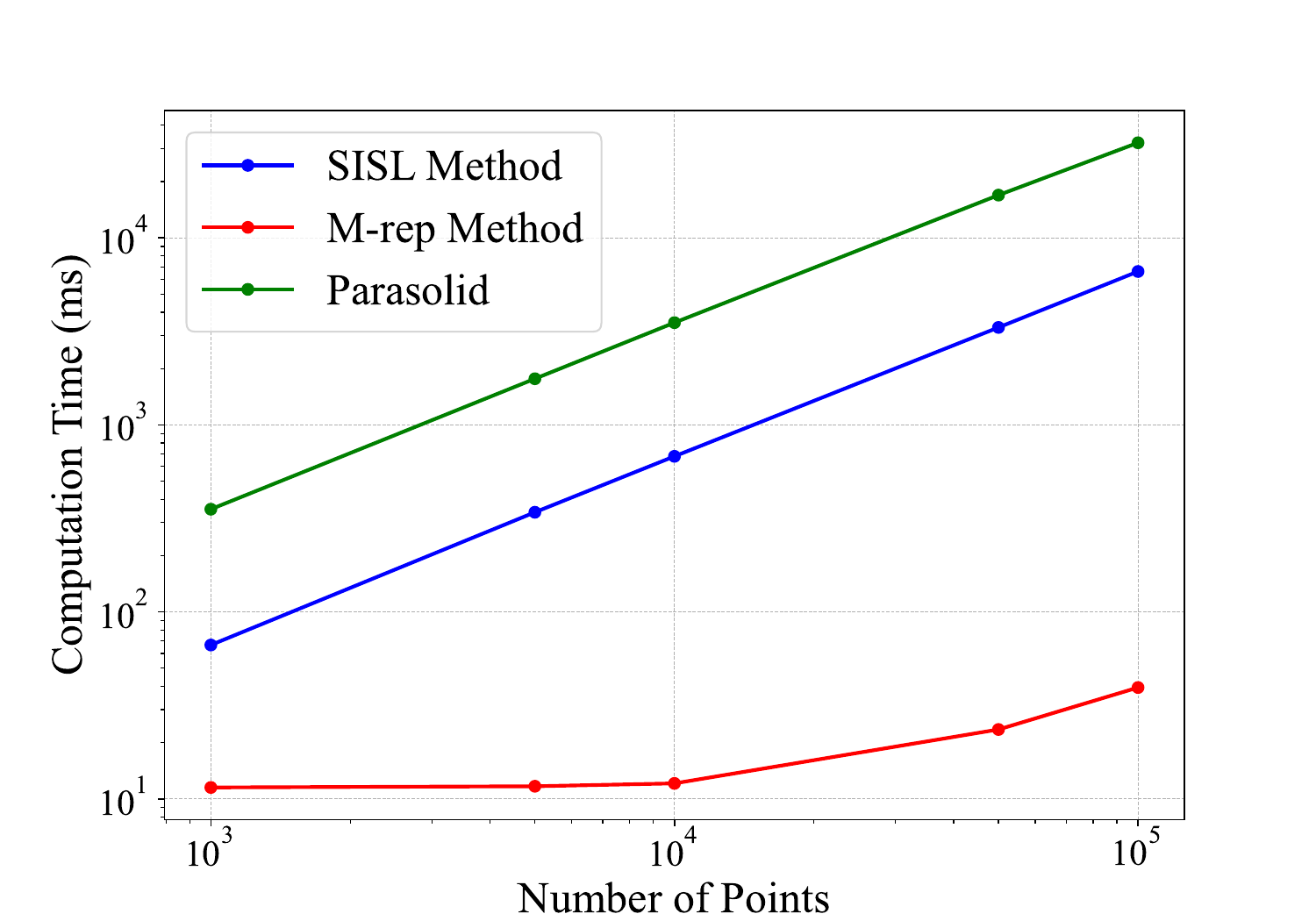}  
        \label{fig:time_comparesion_6hard}
    }
    \caption{Time comparison of M-rep, SISL, and Parasolid on four models from simple to complex.}
    \label{fig:line_plot}
\end{figure}

\begin{figure}[t]
    \centering
    \includegraphics[width=0.5\textwidth]{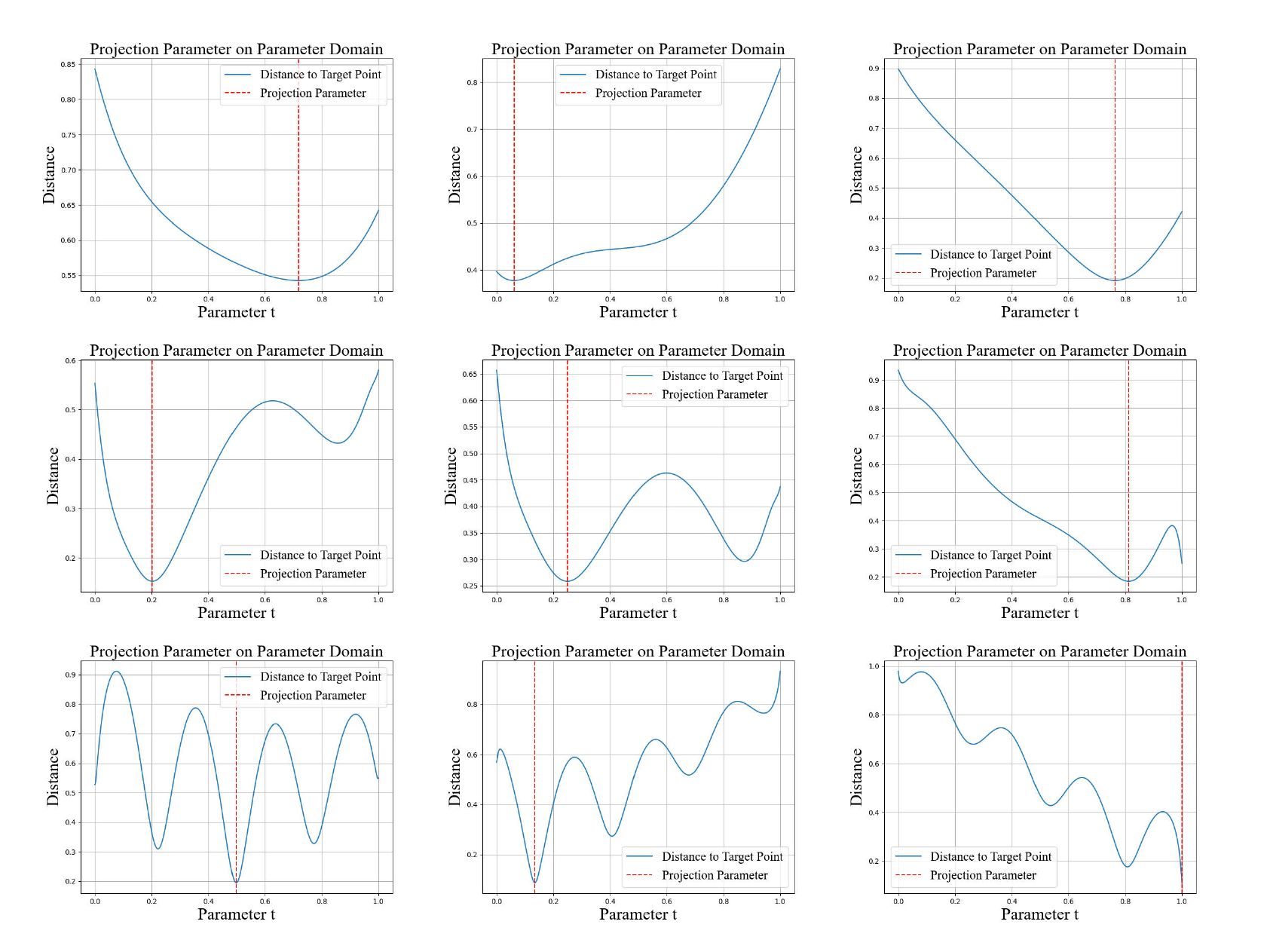}
    \caption{Some projection results based on M-rep.}
    \label{fig:inversion}
\end{figure}

Case studies 1-9 (Fig.~\ref{fig:2D_case}) consider nine 2D B-spline curves of increasing complexity, while case studies 10-18 (Fig.~\ref{fig:3D_case}) focus on nine 3D B-spline curves. All B-spline coordinates are confined within the range of 0 to 1. The performance analysis of each M-rep module is summarized in Fig.~\ref{fig:time_distribution}, Tables~\ref{tab: 2D_case} and~\ref{tab: 3D_case}. We compared the projection and inversion performance across these cases and selected representative examples, highlighting the best, average, and worst performances in Fig.\ref{fig:line_plot}, Tables~\ref{tab: SISL-vs-Parasolid-vs-Mrep_2D} and~\ref{tab: SISL-vs-Parasolid-vs-Mrep_3D}. Table~\ref{tab:inversion} and Fig.~\ref{fig:inversion} demonstrate the robustness of our method. Additional experimental results are available in the supplementary materials.

\begin{table*}[htb]
    \centering
    \small 
    \renewcommand{\arraystretch}{0.8} 
    \setlength\extrarowheight{2pt} 
    \setlength{\abovecaptionskip}{0cm}
    \caption{Projection comparisons of SISL, Parasolid, and M-rep methods.}
    \setlength{\tabcolsep}{4mm} 
    \begin{tabularx}{1.0\textwidth}{
        >{\centering\arraybackslash}>{\hsize=.5\hsize\linewidth=\hsize}X
        >{\centering\arraybackslash}>{\hsize=.5\hsize\linewidth=\hsize}X
        >{\centering\arraybackslash}>{\hsize=.5\hsize\linewidth=\hsize}X
        >{\centering\arraybackslash}>{\hsize=1.5\hsize\linewidth=\hsize}X
        > {\centering\arraybackslash}>{\hsize=0.9\hsize\linewidth=\hsize}X
        > {\centering\arraybackslash}>{\hsize=0.9\hsize\linewidth=\hsize}X
        >{\centering\arraybackslash}>{\hsize=1.7\hsize\linewidth=\hsize}X
        > {\centering\arraybackslash}>{\hsize=0.9\hsize\linewidth=\hsize}X
        > {\centering\arraybackslash}>{\hsize=0.9\hsize\linewidth=\hsize}X
        > {\centering\arraybackslash}>{\hsize=1.7\hsize\linewidth=\hsize}X
    }
    \hline
    \multirow{2}{*}{Model} & \multirow{2}{*}{\makecell[c]{Knots \\Length}} & \multirow{2}{*}{Degree} & \multirow{2}{*}{\makecell[c]{Number of \\Points}} 
    & \multicolumn{3}{c}{Total Time (ms)} & \multicolumn{3}{c}{Average Time ($\mu$s)} \\
    \cline{5-7} \cline{8-10}
    & & & & SISL & Parasolid & M-rep (Ours) & SISL & Parasolid & M-rep (Ours) \\
    \hline
    \multirow{3}{*}{(a)} & \multirow{3}{*}{10} & \multirow{3}{*}{4} & $1 \times 10^{4}$   & 33.3  &  1169  & 2.09  & 3.33  & 117  & 0.209 \\
                             &  &  & $5 \times 10^{4}$ & 164   & 5937 & 3.09  & 3.28  & 119  & 0.062 \\
                             &  &  & $1 \times 10^{5}$  & 303  &  10699  & 3.84  & 3.03  & 107  & 0.038 \\
    \hline
    \multirow{3}{*}{(n)} & \multirow{3}{*}{18} & \multirow{3}{*}{5} & $1 \times 10^{4}$  & 114  &  1750  & 2.34  & 11.4  & 175  & 0.234 \\
                             &  &  & $5 \times 10^{4}$ & 559   & 7735 & 4.61  & 11.2  & 174  & 0.092 \\
                             &  &  & $1 \times 10^{5}$  & 1123  & 14843  & 6.97  & 11.2  & 148  & 0.070 \\
    \hline
    \multirow{3}{*}{(r)} & \multirow{3}{*}{57} & \multirow{3}{*}{6} & $1 \times 10^{4}$  & 679  &  3521  & 12.1  & 67.9  & 352  & 1.21 \\
                             &  &  & $5 \times 10^{4}$ & 3320   & 16935 & 23.5  & 66.4  & 339  & 0.469 \\
                             &  &  & $1 \times 10^{5}$  & 6612  & 32267  & 39.4  & 66.1  & 323  & 0.394 \\
    \hline
    \end{tabularx}
    \label{tab: SISL-vs-Parasolid-vs-Mrep_2D}
\end{table*}

\begin{table*}[htb]
    \centering
    \small 
    \renewcommand{\arraystretch}{0.8} 
    \setlength\extrarowheight{2pt} 
    \setlength{\abovecaptionskip}{0cm}
    \caption{Inversion comparisons of SISL, Parasolid, and M-rep methods.}
    \setlength{\tabcolsep}{4mm} 
    \begin{tabularx}{1.0\textwidth}{
        >{\centering\arraybackslash}>{\hsize=.5\hsize\linewidth=\hsize}X
        >{\centering\arraybackslash}>{\hsize=.5\hsize\linewidth=\hsize}X
        >{\centering\arraybackslash}>{\hsize=.5\hsize\linewidth=\hsize}X
        >{\centering\arraybackslash}>{\hsize=1.5\hsize\linewidth=\hsize}X
        > {\centering\arraybackslash}>{\hsize=0.9\hsize\linewidth=\hsize}X
        > {\centering\arraybackslash}>{\hsize=0.9\hsize\linewidth=\hsize}X
        >{\centering\arraybackslash}>{\hsize=1.7\hsize\linewidth=\hsize}X
        > {\centering\arraybackslash}>{\hsize=0.9\hsize\linewidth=\hsize}X
        > {\centering\arraybackslash}>{\hsize=0.9\hsize\linewidth=\hsize}X
        > {\centering\arraybackslash}>{\hsize=1.7\hsize\linewidth=\hsize}X
    }
    \hline
    \multirow{2}{*}{Model} & \multirow{2}{*}{\makecell[c]{Knots \\Length}} & \multirow{2}{*}{Degree} & \multirow{2}{*}{\makecell[c]{Number of \\Points}} 
    & \multicolumn{3}{c}{Total Time (ms)} & \multicolumn{3}{c}{Average Time ($\mu$s)} \\
    \cline{5-7} \cline{8-10}
    & & & & SISL & Parasolid & M-rep (Ours) & SISL & Parasolid & M-rep (Ours) \\
    \hline
    \multirow{3}{*}{(l)} & \multirow{3}{*}{14} & \multirow{3}{*}{6} & $1 \times 10^{4}$   & 69.0  &  1684  & 12.3  & 6.90  & 168  & 1.23 \\
                             &  &  & $5 \times 10^{4}$ & 343   & 7293 & 17.1  & 6.86  & 146  & 0.341 \\
                             &  &  & $1 \times 10^{5}$  & 704  & 13991  & 24.8  & 7.04  & 140  & 0.248 \\
    \hline
    \multirow{3}{*}{(o)} & \multirow{3}{*}{22} & \multirow{3}{*}{6} & $1 \times 10^{4}$  & 184  &  1993  & 8.47  & 18.4  & 200  & 0.847 \\
                             &  &  & $5 \times 10^{4}$ & 876   & 8799 & 11.6  & 17.5  & 176  & 0.232 \\
                             &  &  & $1 \times 10^{5}$  & 1753  & 17021  & 16.1  & 17.5  & 170  & 0.161 \\
    \hline
    \multirow{3}{*}{(q)} & \multirow{3}{*}{46} & \multirow{3}{*}{5} & $1 \times 10^{4}$   & 387  &  2988  & 5.00  & 38.7  & 299  & 0.500 \\
                             &  &  & $5 \times 10^{4}$ & 1973   & 13137 & 15.8  & 39.5  & 263  & 0.316 \\
                             &  &  & $1 \times 10^{4}$  & 3845  & 26924  & 27.3  & 38.5  & 269  & 0.273 \\
    \hline
    \end{tabularx}
    \label{tab: SISL-vs-Parasolid-vs-Mrep_3D}
\end{table*}

\begin{table}[t]
\caption{Inversion error of M-rep on Fig.~\ref{fig:case14}.}
\centering
\resizebox{\columnwidth}{!}{ 
\begin{tabular}{ccc}
\hline
Point & Approximated Parameter & Error \\
\hline
(0.000000, 0.100000, 0.200000) & 0.000000 & 0.000000 \\
(1.000000, 0.800000, 0.500000) & 1.000000 & 0.000000 \\
(0.438773, 0.538773, 0.838753) & 0.361290 & 0.000017 \\
(0.577418, 0.677418, 0.948643) & 0.564589 & 0.000003 \\
(0.715229, 0.815229, 0.883098) & 0.749884 & 0.000000 \\
(0.660546, 0.760546, 0.928540) & 0.681020 & 0.000001 \\
\hline
\end{tabular}
}
\label{tab:inversion}
\end{table}

\subsection{Discussion and limitation}
As illustrated in Fig.~\ref{fig:time_distribution} and summarized in Tables~\ref{tab: 2D_case} and~\ref{tab: 3D_case}, the majority of the computation time is spent on the decomposition stage. This is because performing accurate calculations in subsequent steps requires extensive approximation of the original B-splines. Additionally, as the number of test points increases, the time needed for the error-controlled decomposition step basically remains unchanged. This trend is expected, as error-controlled decomposition is a preprocessing step, its performance is independent of the number of points.

Tables~\ref{tab: SISL-vs-Parasolid-vs-Mrep_2D} and~\ref{tab: SISL-vs-Parasolid-vs-Mrep_3D} show that M-rep can handle approximately 10 million projection and inversion tasks per second for large-scale tasks in most cases, offering an improvement of nearly two orders of magnitude compared to SISL and Parasolid, as further visualized in Fig.~\ref{fig:line_plot}.
These results highlight the significant performance improvement that M-rep brings to B-spline computations.

Beyond performance, our method also demonstrates robustness. Fig.~\ref{fig:inversion} demonstrates that M-rep accurately identifies the parameter value with the minimum distance in both simple and complex scenarios. As shown in Table ~\ref{tab:inversion}, the error in our inversion results can be maintained within the order of $1\mathrm{e}{-4}$.

It is important to note that due to the typical GPU block size of 1024, when the degree of the curve exceeds 31, a single block is insufficient to compute the polynomial coefficient matrix of the B-spline basis function. As a result, the performance of our pre-calculation process will be impacted. Addressing the challenge of maintaining high performance with ultra-high-degree B-splines will be an important area for future work.

\section{Conclusions}
\label{sec:conclusion}
This paper extends matrix-based GPU computation to advanced B-spline operations, particularly for high-degree B-splines with varying knots and control points. The main feature of this method is its high efficiency, which can achieve an approximately two-order-of-magnitude speedup over existing methods for projection and inversion operations. This feature is essentially achieved by decomposing B-splines into cubic Béziers under an error-controlled mechanism, which ensures parameter uniformity and therefore facilitates GPU processing, as well as by resolving the challenges in warp and memory divergence to optimize GPU computation, achieving high performance. The empirical validation of the proposed method’s effectiveness has been demonstrated using a series of examples and comparisons.

Despite the promising results, the method has certain limitations. Most notably, the current idea of using GPUs to accelerate CAD modeling is limited to low-level operations such as knot insertion, projection, and inversion, without addressing higher-level functionalities like direct edits~\cite{zou2023variational,zou2019push}, parametric edits~\cite{tang2023decision}, Boolean operations~\cite{requicha1985boolean}, or extending into CAM applications such as toolpath generation~\cite{zou2013iso,zou2021length,zou2021robust} and machining simulation~\cite{gong2016cutter}. Expanding the scope of the method to encompass these areas is one of the key directions for future research.
It is currently designed for non-rational B-splines, with rational B-splines requiring additional processing steps for accurate computation, which limits its applicability in scenarios involving rational B-splines. Extending the approach to efficiently handle rational B-splines represents a key direction for future research. 
Moreover, when dealing with B-splines of degrees higher than 31, the GPU block size becomes a limiting factor, making it insufficient to vectorize the B-spline basis function within a single block. Addressing the challenge will be a critical direction for future research, with a focus on optimizing workload distribution and GPU resource utilization.

\section*{Acknowledgements}
This work has been funded by NSF of China (No. 62102355) and the ``Pioneer" and ``Leading Goose" R\&D Program of Zhejiang Province (No. 2024C01103).

\bibliographystyle{elsarticle-num} 
\bibliography{cas-refs}






\section*{Appendix}
\subsection{Mathematical derivation of B-spline matrix decomposition}
\label{decomposition derivation}
For a B-spline curve $C(t)$ within the interval \( t \in [t_q, t_{q+1}] \), we can rewrite it in matrix representation:
\begin{equation}
C(t) = \begin{bmatrix} 1 & t & t^2 & \cdots & t^{p} \end{bmatrix} 
A_{q,p,T} 
\begin{bmatrix} 
P_{q-p} \\
P_{q-p+1} \\
\vdots \\
P_q
\end{bmatrix}
\label{eq:bsplinematrix}
\end{equation}
after decomposition, Eq.~\eqref{eq:bsplinematrix} can be written as:
\begin{equation}
C(t) = R(u) = \begin{bmatrix} 1 & u & u^2 & \cdots & u^{p} \end{bmatrix} 
B_{p} 
\begin{bmatrix} 
Q_{q,0} \\
Q_{q,1} \\
\vdots \\
Q_{q,p}
\end{bmatrix}
\label{eq:bsplineafter}
\end{equation}
the parameter range \( t \in [t_q, t_{q+1}] \) normalized to a local parameter \( u \in [0, 1] \) is as follows:
\[
u = \frac{t - t_q}{t_{q+1} - t_q}, \quad t = u (t_{q+1} - t_q) + t_q.
\]
the transformation matrix \( M_{q,p,T}  \) is defined such that:
\begin{equation}
T = U M_{q,p,T} 
\label{eq:appMqpt}
\end{equation}
where \( T = [1, t, t^2, \dots, t^p]^\top \), \( U = [1, u, u^2, \dots, u^p]^\top \), using the binomial theorem:
\begin{equation}
t^i = \left( u (t_{q+1} - t_q) + t_q \right)^i = \sum_{k=0}^i \binom{i}{k} (t_{q+1} - t_q)^k u^k t_q^{i-k}
\label{appbinomial theorem}
\end{equation}we can get the entries of \( M_{q,p,T}  \) are:
\begin{equation}
m_{ji} = 
\begin{cases} 
\binom{i}{j} (t_{q+1} - t_q)^j t_q^{i-j}, & \text{if } j \leq i \\
0, & \text{if } j > i
\end{cases}
\label{eq:appMelement}
\end{equation}
through Eq.~\eqref{eq:bsplinematrix} and Eq.~\eqref{eq:bsplineafter}, we can get:
\begin{equation}
T  
A_{q,p,T} 
\begin{bmatrix} 
P_{q-p} \\
P_{q-p+1} \\
\vdots \\
P_q
\end{bmatrix} = U 
B_{p} 
\begin{bmatrix} 
Q_{q,0} \\
Q_{q,1} \\
\vdots \\
Q_{q,p}
\end{bmatrix}
\label{eq:appbspline=bezier}
\end{equation}
by substituting Eq.~\eqref{eq:appMqpt} into Eq.~\eqref{eq:appbspline=bezier}, we get:
\begin{equation}
U M_{q,p,T}   
A_{q,p,T} 
\begin{bmatrix} 
P_{q-p+1} \\
P_{q-p+2} \\
\vdots \\
P_q
\end{bmatrix} = U
B_{p} 
\begin{bmatrix} 
Q_{q,0} \\
Q_{q,1} \\
\vdots \\
Q_{q,p}
\end{bmatrix}
\end{equation}
by shifting the terms on the right side of the equation we can get we can get:
\begin{equation}
U (M_{q,p,T}  
A_{q,p,T} 
\begin{bmatrix} 
P_{q-p+1} \\
P_{q-p+2} \\
\vdots \\
P_q
\end{bmatrix} - 
B_{p} 
\begin{bmatrix} 
Q_{q-p+1} \\
Q_{q-p+2} \\
\vdots \\
Q_q
\end{bmatrix} ) = 0
\label{eq:app6}
\end{equation}
since Eq.~\eqref{eq:app6} holds for all $U$, so: 
\begin{equation}
M_{q,p,T} 
A_{q,p,T} 
\begin{bmatrix} 
P_{q-p+1} \\
P_{q-p+2} \\
\vdots \\
P_q
\end{bmatrix} = 
B_{p} 
\begin{bmatrix} 
Q_{q-p+1} \\
Q_{q-p+2} \\
\vdots \\
Q_q
\end{bmatrix}
\end{equation}
because for any $p$, the inverse of $B_{p}$ always exists, so:
\begin{equation}
\begin{bmatrix} 
Q_{q,0} \\
Q_{q,1} \\
\vdots \\
Q_{q,p}
\end{bmatrix} = B_{p}^{-1} M_{q,p,T} 
A_{q,p,T} 
\begin{bmatrix} 
P_{q-p+1} \\
P_{q-p+2} \\
\vdots \\
P_q
\end{bmatrix}
\label{eq:appdecomposition}
\end{equation}
define :
\begin{equation}
D_{q,p} = M_{q,p,T} 
A_{q,p,T} 
\begin{bmatrix} 
P_{q-p+1} \\
P_{q-p+2} \\
\vdots \\
P_q
\end{bmatrix}
\label{Dqp}
\end{equation}
substituting Eq.~\eqref{Dqp} into Eq.~\eqref{eq:appdecomposition}, we get
\begin{equation}
\begin{bmatrix} 
Q_{q,0} \\
Q_{q,1} \\
\vdots \\
Q_{q,p}
\end{bmatrix} = B_{p}^{-1} D_{q,p}
\label{eq:appdecomfinal}
\end{equation}
where $\begin{bmatrix} 
Q_{q,0} \\
Q_{q,1} \\
\vdots \\
Q_{q,p}
\end{bmatrix}$ is the control points matrix of the Bezier segment decomposed from the $q$-th knot interval.

\subsection{Proof that reparametrizing $T$ is more efficient than reparametrizing $U$ in M-rep for B-spline decomposition.}

During the B-spline decomposition process of many methods, we typically choose to transform the parameters of $U$, in which case, the transformation matrix \( M_{q,p,U}  \) is defined such that:
\begin{equation}
U = T M_{q,p,U} 
\label{eq:appMqpu}
\end{equation}
According to Sec.~\ref{decomposition derivation}, we can get:
\begin{equation}
T    
A_{q,p,T} 
\begin{bmatrix} 
P_{q-p+1} \\
P_{q-p+2} \\
\vdots \\
P_q
\end{bmatrix} = T M_{q,p,U}
B_{p} 
\begin{bmatrix} 
Q_{q,0} \\
Q_{q,1} \\
\vdots \\
Q_{q,p}
\end{bmatrix}
\end{equation}
doing similar calculations as in Sec.~\ref{decomposition derivation}, we can finally get
\begin{equation}
\begin{bmatrix} 
Q_{q,0} \\
Q_{q,1} \\
\vdots \\
Q_{q,p}
\end{bmatrix} = \left( M_{q,p,U} B_{p} \right)^{-1} A_{q,p,T} 
\begin{bmatrix} 
P_{q-p+1} \\
P_{q-p+2} \\
\vdots \\
P_q
\end{bmatrix}
\end{equation}

In this case, while tensor cores can be used to calculate $M_{q,p,U}$ and $B_p$, an additional inversion operation is required on the result of their multiplication. However, inverting large matrices is both complex and time-consuming. So it is obviously not as efficient as Eq.~\eqref{eq:appdecomfinal}.

\subsection{Mathematical derivation of $L_2$-error minimized $G_1$-continuous reduction}
For the $p$-degree Bezier curve with control points $\mathbf{Q}=\begin{bmatrix}
Q_{0}, Q_{1} \dots Q_{p}
\end{bmatrix}^T$: 
\[
Q(t) = \sum_{i=0}^{p} B_{i,p}(t) Q_{i} =: B_{p} \mathbf{Q}, \quad 0 \leq t \leq 1,
\]
we need to find an approximated cubic Bézier curve
\[
R(t) = \sum_{i=0}^{3} B_{i,3}(t) R_{i} =: B_{3} \mathbf{R}, \quad 0 \leq t \leq 1, 
\]
of lower degree $3$, with the control points $\mathbf{R}=\begin{bmatrix}
R_{0}, R_{1} \dots R_{3}
\end{bmatrix}^T$, so that the following two conditions are satisfied:
\begin{enumerate}
    \item[(i)] $Q(t)$ and $R(t)$ are $G^1$-continuous at the end points.
    \item[(ii)] the $L_2$-error $\epsilon$ between $Q(t)$ and $R(t)$ is minimum.
\end{enumerate}

For $G^1$-continuous degree reduction constraint, the following conditions must be satisfied:
\begin{equation}
\begin{aligned}
R_{0} = Q_{0} &  
\quad R_{1} = Q_{0} + \frac{p}{3}  \Delta Q_{0}\delta_0 \\
R_{3} = Q_{p} & 
\quad R_{2} = Q_{p} - \frac{p}{3}  \Delta Q_{p-1}\delta_1
\end{aligned}
\label{appG1-condition}
\end{equation}
where $\Delta Q_{i}$ means $Q_{i+1} - Q_{i}$, $\delta_0$ and $\delta_1$ are two variables.
And minimizing $L_2$-error means we need to minimize:
\begin{equation}
\varepsilon = \int_{0}^{1} \|B_p Q - B_3 R\|^2 \, dt
\label{appepsl}
\end{equation}
The minimum occurs when the partial derivatives vanish. Differentiating it with respect to $\delta_i$ and equating to zero gives
\begin{equation}
\frac{\partial \varepsilon}{\partial \delta_0} = 
\left( G_{3,p}^1 Q - G_{3,3}^{1} R \right) \cdot \Delta Q_{0} = 0
\label{appdifferentiatedelta0}\end{equation}
\begin{equation}
\frac{\partial \varepsilon}{\partial \delta_1} = 
\left( G_{3,p}^{2} Q - G_{3,3}^{2} R \right) \cdot \Delta Q_{p-1} = 0
\label{appdifferentiatedelta1}
\end{equation}
where \( G_{m,n}^{i} \) is the submatrix of \( G_{m,n} \) formed by the $i$-th rows.
Substitute Eq.~\eqref{appG1-condition} into Eq.~\eqref{appdifferentiatedelta0} and Eq.~\eqref{appdifferentiatedelta1}, we get:
\begin{equation}
\left( G_{3,p}^1 Q - G_{3,3}^1 \left[ Q_0, Q_0 + \frac{p}{3} \Delta Q_0\delta_0, Q_{p} - \frac{p}{3}  \Delta Q_{p-1}\delta_1, Q_{p} \right] \right) \cdot \Delta Q_0 = 0
\label{B5}
\end{equation}
\begin{equation}
\left( G_{3,p}^2 Q - G_{3,3}^2 \left[ Q_0, Q_0 + \frac{p}{3} \Delta Q_0\delta_0, Q_{p} - \frac{p}{3}  \Delta Q_{p-1}\delta_1, Q_{p} \right] \right) \cdot \Delta Q_{p-1} = 0
\label{B6}
\end{equation}
set
\begin{equation}
V_1 = G_{3,p}^1 Q - G_{3,3}^1 [Q_0, Q_0, Q_p, Q_p]
\label{V1}
\end{equation}
\begin{equation}
V_2 = G_{3,p}^2 Q - G_{3,3}^2 [Q_0, Q_0, Q_p, Q_p]
\label{V2}
\end{equation}
substitute Eq.~\eqref{V1} and Eq.~\eqref{V2} into Eq.~\eqref{B5} and Eq.~\eqref{B6}, we get
\begin{equation}
\left( V_1 - G_{3,3}^1 \left[ 0, \frac{p}{3} \Delta Q_0 \delta_0, -\frac{p}{3} \Delta Q_{p-1} \delta_1, 0 \right] \right) \cdot \Delta Q_0 = 0
\end{equation}

\begin{equation}
\left( V_2 - G_{3,3}^2 \left[ 0, \frac{p}{3} \Delta Q_0 \delta_0, -\frac{p}{3} \Delta Q_{p-1} \delta_1, 0 \right] \right) \cdot \Delta Q_{p-1} = 0
\end{equation}
Next we define the quantities after linear calculation:
\[
A_0 = G_{3,3}^1 \left[ 0, \frac{p}{3} \Delta Q_0, 0, 0 \right], \quad A_1 = G_{3,3}^1 \left[ 0, 0, -\frac{p}{3} \Delta Q_{p-1}, 0 \right],
\]
\[
B_0 = G_{3,3}^2 \left[ 0, \frac{p}{3} \Delta Q_0, 0, 0 \right], \quad B_1 = G_{3,3}^2 \left[ 0, 0, -\frac{p}{3} \Delta Q_{p-1}, 0 \right].
\]
The equation can be transformed into: 
\begin{equation}
V_1 \cdot \Delta Q_0 - \delta_0 (A_0 \cdot \Delta Q_0) + \delta_1 (A_1 \cdot \Delta Q_0) = 0
\label{V1equation}
\end{equation}
\begin{equation}
V_2 \cdot \Delta Q_{p-1} - \delta_0 (B_0 \cdot \Delta Q_{p-1}) + \delta_1 (B_1 \cdot \Delta Q_{p-1}) = 0
\label{V2equation}
\end{equation}
then we write Eq.~\eqref{V1equation} and Eq.~\eqref{V2equation} in matrix form:
\begin{equation}
\begin{cases}
a_{11} \delta_0 + a_{12} \delta_1 = b_1 \\
a_{21} \delta_0 + a_{22} \delta_1 = b_2
\end{cases}
\label{Vmatrix}
\end{equation}
where
\[
a_{11} = - A_0 \cdot \Delta Q_0, \quad a_{12} = A_1 \cdot \Delta Q_0, \quad b_1 = - V_1 \cdot \Delta Q_0,
\]
\[
a_{21} = - B_0 \cdot \Delta Q_{p-1}, \quad a_{22} = B_1 \cdot \Delta Q_{p-1}, \quad b_2 = - V_2 \cdot \Delta Q_{p-1}.
\]
by using Cramer's rule, we can find the determinant of the Eq.~\eqref{Vmatrix} as:
\begin{equation}
D = a_{11} a_{22} - a_{12} a_{21}
\label{determinant}
\end{equation}
therefore, we can get:
\[
\delta_0 = \frac{b_1 a_{22} - a_{12} b_2}{D}, \quad \delta_1 = \frac{a_{11} b_2 - b_1 a_{21}}{D}.
\]
So far, we can get the final solution:
\begin{equation}
\delta_0 = \frac{(V_2 \cdot \Delta Q_{p-1}) (A_1 \cdot \Delta Q_0) - (V_1 \cdot \Delta Q_0) (B_1 \cdot \Delta Q_{p-1})}{(A_0 \cdot \Delta Q_0)(B_1 \cdot \Delta Q_{p-1}) - (A_1 \cdot \Delta Q_0)(B_0 \cdot \Delta Q_{p-1})}
\label{delta0value}
\end{equation}
\begin{equation}
\delta_1 = \frac{(A_0 \cdot \Delta Q_0) (V_2 \cdot \Delta Q_{p-1}) - (B_0 \cdot \Delta Q_{p-1}) (V_1 \cdot \Delta Q_0)}{(A_0 \cdot \Delta Q_0)(B_1 \cdot \Delta Q_{p-1}) - (A_1 \cdot \Delta Q_0)(B_0 \cdot \Delta Q_{p-1})}
\label{delta1value}
\end{equation}

\subsection{Detailed discussion of tensor core accelerated power to Bernstein transformation}
Each Bernstein basis polynomial $B_{j,n}(t) = \binom{n}{j} t^j (1 - t)^{n-j}$
 can be expanded into a power basis form:
\[
B_{j,n}(t) = \sum_{k=j}^{n} \binom{n}{j} \binom{n-j}{k-j} (-1)^{k-j} t^k
\]
this expansion shows that the elements of the transformation matrix $M$ from the Bernstein basis to the power basis ($\mathbf{a} = M \mathbf{b}$) are:
\[
M_{k,j} = 
\begin{cases} 
\binom{n}{j} \binom{n-j}{k-j} (-1)^{k-j} & \text{if } k \geq j, \\
0 & \text{otherwise}.
\end{cases}
\]
by calculating $T = M^{-1}$, we can get:
\[
T_{i,j} =
\begin{cases}
\binom{i}{j} \cdot \binom{n}{i}^{-1}, & \text{if } j \leq i, \\
0, & \text{otherwise}.
\end{cases}
\]

\begin{figure}
    \centering
    \includegraphics[width=0.5\linewidth]{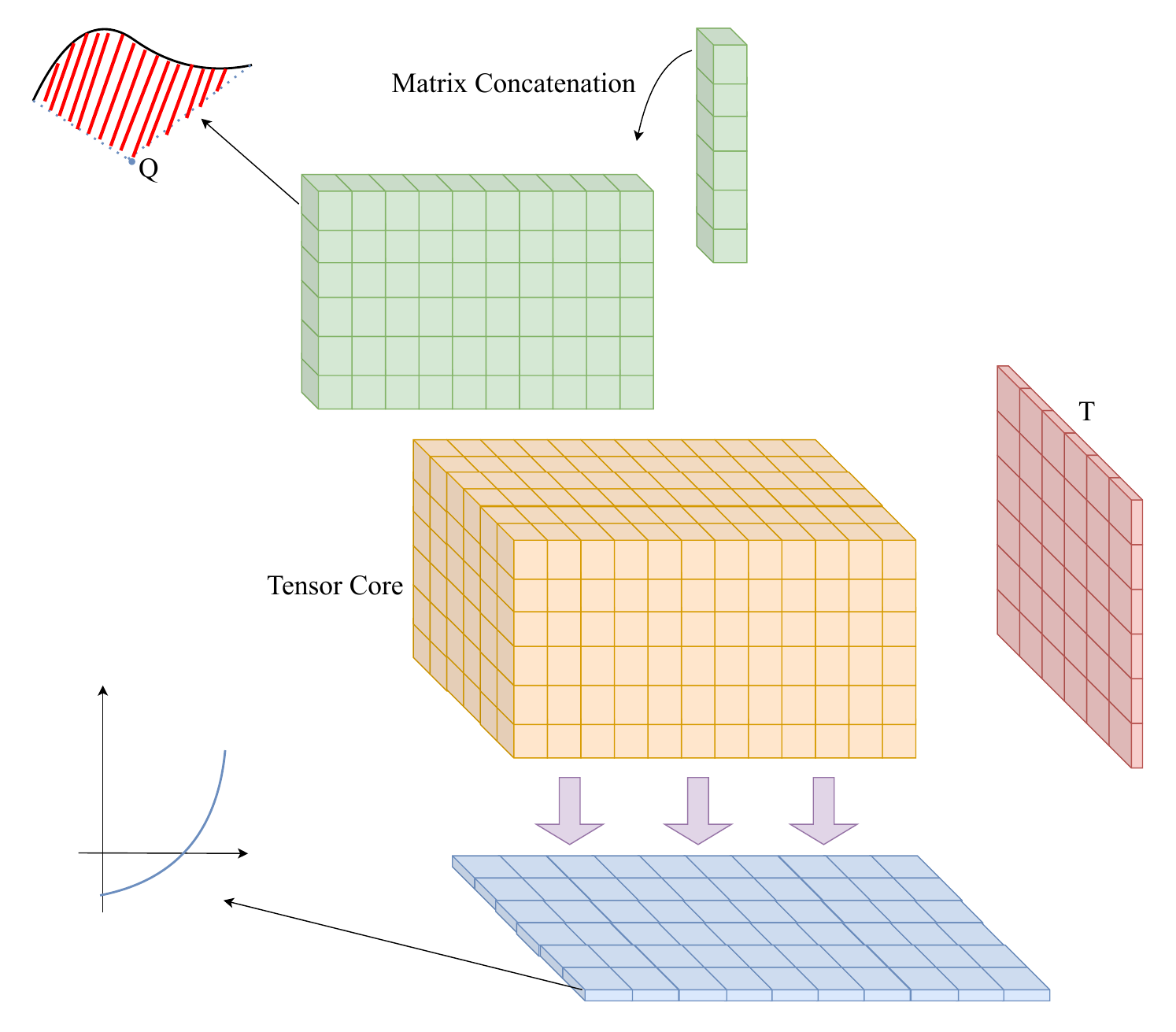}
    \caption{Pipeline of tensor core accelerated geometric conversion.}
    \label{fig:geometric conversion}
\end{figure}

The illustration of tensor core accelerated power basis to Bernstein basis conversion can be seen in Fig.~\ref{fig:geometric conversion}. Each $E(t)$ is represented as a $1 \times 6$ vector of polynomial coefficients. These vectors can be concatenated into a large matrix, which is then multiplied by T using the tensor core to obtain the coefficients of $E(t)$ in the Bernstein basis.

\subsection{Additional Experiments}
\begin{figure}
    \centering
    \includegraphics[width=1.0\linewidth]{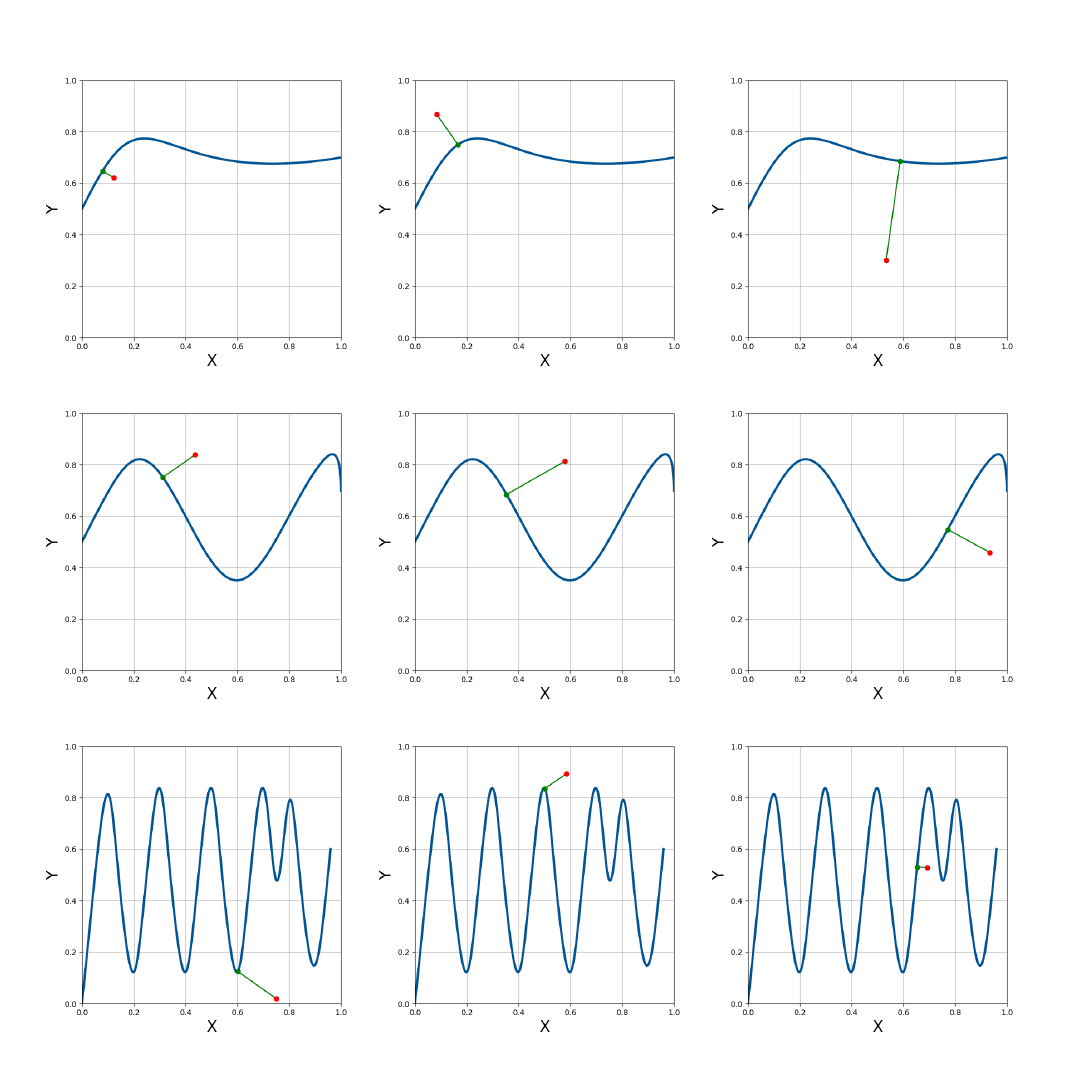}
    \caption{Some projection results based on M-rep on 2D curves.}
    \label{fig:2D-result}
\end{figure}

\begin{figure}
    \centering
    \includegraphics[width=1.0\linewidth]{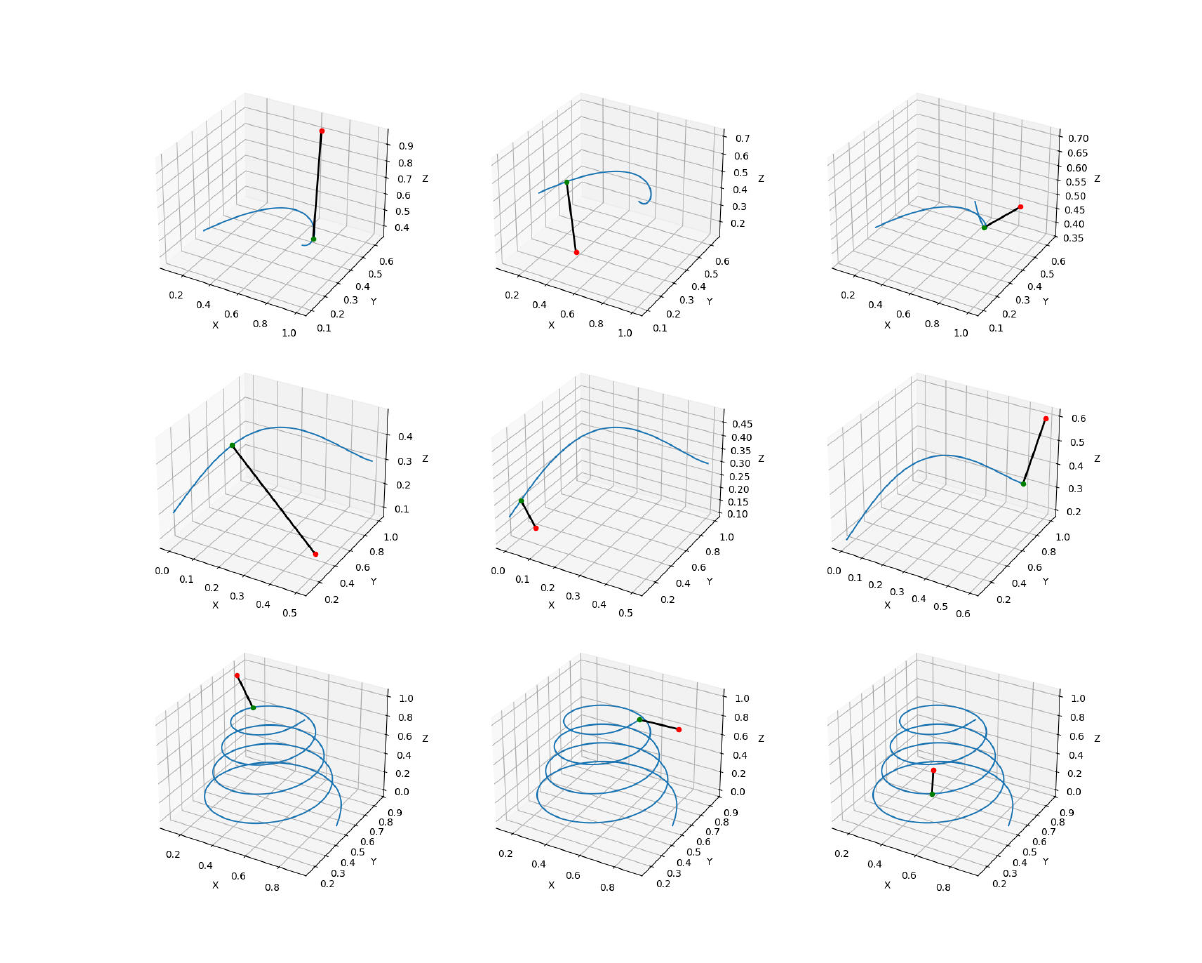}
    \caption{Some projection results based on M-rep on 3D curves.}
    \label{fig:3D-result}
\end{figure}

Our complete comparison can be seen in Tables~\ref{tab: SISL-vs-Parasolid-vs-Mrep_2D},~\ref{tab: SISL-vs-Parasolid-vs-Mrep_3D} and~\ref{tab: SISL-vs-Parasolid-vs-Mrep_3D_inversion}. Additional projection results can be seen in Fig.~\ref{fig:2D-result} and Fig.~\ref{fig:3D-result}. And performace comparison between numerical and analytical methods for monotonic decomposition is shown in Table~\ref{tab:analytical}.

\begin{table*}[htb]
    \centering
    \small 
    \renewcommand{\arraystretch}{0.8} 
    \setlength\extrarowheight{2pt} 
    \setlength{\abovecaptionskip}{0cm}
    \caption{Projection Comparisons of SISL, Parasolid, and M-rep method on 2D cases.}
    \setlength{\tabcolsep}{4mm} 
    \begin{tabularx}{1.0\textwidth}{
        >{\centering\arraybackslash}>{\hsize=.5\hsize\linewidth=\hsize}X
        >{\centering\arraybackslash}>{\hsize=.5\hsize\linewidth=\hsize}X
        >{\centering\arraybackslash}>{\hsize=.5\hsize\linewidth=\hsize}X
        >{\centering\arraybackslash}>{\hsize=1.5\hsize\linewidth=\hsize}X
        >{\centering\arraybackslash}X
        >{\centering\arraybackslash}X
        >{\centering\arraybackslash}X
        >{\centering\arraybackslash}X
        >{\centering\arraybackslash}X
        >{\centering\arraybackslash}X
    }
    \hline
    \multirow{2}{*}{Model} & \multirow{2}{*}{\makecell[c]{Knots \\Length}} & \multirow{2}{*}{Degree} & \multirow{2}{*}{\makecell[c]{Number of \\Points}} 
    & \multicolumn{3}{c}{Total Time (ms)} & \multicolumn{3}{c}{Average Time ($\mu$s)} \\
    \cline{5-7} \cline{8-10}
    & & & & SISL & Parasolid & M-rep (ours) & SISL & Parasolid & M-rep (ours)\\
    \hline
    \multirow{3}{*}{(a)} & \multirow{3}{*}{10} & \multirow{3}{*}{4} & $1 \times 10^{4}$  & 33.3  &  1169  & 2.09  & 3.33  & 117  & 0.209 \\
                             &  &  & $5 \times 10^{4}$ & 164   & 5937 & 3.09  & 3.28  & 119  & 0.062 \\
                             &  &  & $1 \times 10^{5}$  & 303  &  10699  & 3.84  & 3.03  & 107  & 0.038 \\
    \hline
    \multirow{3}{*}{(b)} & \multirow{3}{*}{12} & \multirow{3}{*}{5} & $1 \times 10^{4}$  & 35.6  & 1196   & 9.51  & 3.56  & 120  & 0.951 \\
                             &  &  & $5 \times 10^{4}$ & 185   & 6073  & 11.0  & 3.70  & 121  & 0.219 \\
                             &  &  & $1 \times 10^{5}$  & 347  & 12504   & 17.1  & 3.47  & 125  & 0.171 \\
    \hline
    \multirow{3}{*}{(c)} & \multirow{3}{*}{14} & \multirow{3}{*}{6} & $1 \times 10^{4}$  & 62.7   & 1764    & 12.2  & 6.27   & 176  & 1.221 \\
                             &  &  & $5 \times 10^{4}$ & 237   & 7203  & 19.1  & 4.73  & 144  & 0.381 \\
                             &  &  & $1 \times 10^{5}$  & 451  & 14168   & 26.1  & 4.51  & 142  & 0.261 \\
    \hline
    \multirow{3}{*}{(d)} & \multirow{3}{*}{14} & \multirow{3}{*}{4} & $1 \times 10^{4}$  & 59.9  &  1705  & 5.67  & 5.99  & 171  & 0.567 \\
                             &  &  & $5 \times 10^{4}$ & 287   & 7679 & 8.43  & 5.73  & 154  & 0.168 \\
                             &  &  & $1 \times 10^{5}$  & 580  &  15242  & 14.6  & 5.80  & 152  & 0.146 \\
    \hline
    \multirow{3}{*}{(e)} & \multirow{3}{*}{18} & \multirow{3}{*}{5} & $1 \times 10^{4}$  & 129  & 2196   & 2.72  & 12.9  & 220  & 0.272 \\
                             &  &  & $5 \times 10^{4}$ & 574   & 9214  & 5.12  & 11.5  & 184  & 0.102 \\
                             &  &  & $1 \times 10^{5}$  & 1127  & 17988   & 8.34  & 11.3  & 180  & 0.083 \\
    \hline
    \multirow{3}{*}{(f)} & \multirow{3}{*}{22} & \multirow{3}{*}{6} & $1 \times 10^{4}$  & 196   & 2319    & 7.85  & 19.6   & 232  & 0.785 \\
                             &  &  & $5 \times 10^{4}$ & 898   & 9968  & 13.0  & 18.0  & 199  & 0.260 \\
                             &  &  & $1 \times 10^{5}$  & 1803  & 19819   & 15.4  & 18.0  & 198  & 0.154 \\
    \hline
    \multirow{3}{*}{(g)} & \multirow{3}{*}{35} & \multirow{3}{*}{4} & $1 \times 10^{4}$  & 210  &  3423  & 5.32  & 21.0  & 342  & 0.532 \\
                             &  &  & $5 \times 10^{4}$ & 1036   & 16911 & 14.9  & 20.7  & 338  & 0.298 \\
                             &  &  & $1 \times 10^{5}$  & 2113  &  33368  & 25.4  & 21.1  & 334  & 0.254 \\
    \hline
    \multirow{3}{*}{(h)} & \multirow{3}{*}{22} & \multirow{3}{*}{5} & $1 \times 10^{4}$  & 139  & 2190   & 3.37  & 13.9  & 219  & 0.337 \\
                             &  &  & $5 \times 10^{4}$ & 690   & 9980  & 6.74  & 13.8  & 200  & 0.135 \\
                             &  &  & $1 \times 10^{5}$  & 1361  & 18016   & 10.8  & 13.6  & 180  & 0.108 \\
    \hline
    \multirow{3}{*}{(i)} & \multirow{3}{*}{57} & \multirow{3}{*}{6} & $1 \times 10^{4}$  & 672   & 3248    & 9.85  & 67.2   & 325  & 0.985 \\
                             &  &  & $5 \times 10^{4}$ & 3310   & 15856  & 25.8  & 66.2  & 317  & 0.517 \\
                             &  &  & $1 \times 10^{5}$  & 6699  & 31782   & 37.0  & 67.0  & 318  & 0.370 \\
    \hline
    \end{tabularx}
    \label{tab: SISL-vs-Parasolid-vs-Mrep_2D}
\end{table*}

\begin{table*}[htb]
    \centering
    \small 
    \renewcommand{\arraystretch}{0.8} 
    \setlength\extrarowheight{2pt} 
    \setlength{\abovecaptionskip}{0cm}
    \caption{Projection Comparisons of SISL, Parasolid, and M-rep method on 3D cases.}
    \setlength{\tabcolsep}{4mm} 
    \begin{tabularx}{1.0\textwidth}{
        >{\centering\arraybackslash}>{\hsize=.5\hsize\linewidth=\hsize}X
        >{\centering\arraybackslash}>{\hsize=.5\hsize\linewidth=\hsize}X
        >{\centering\arraybackslash}>{\hsize=.5\hsize\linewidth=\hsize}X
        >{\centering\arraybackslash}>{\hsize=1.5\hsize\linewidth=\hsize}X
        >{\centering\arraybackslash}X
        >{\centering\arraybackslash}X
        >{\centering\arraybackslash}X
        >{\centering\arraybackslash}X
        >{\centering\arraybackslash}X
        >{\centering\arraybackslash}X
    }
    \hline
    \multirow{2}{*}{Model} & \multirow{2}{*}{\makecell[c]{Knots \\Length}} & \multirow{2}{*}{Degree} & \multirow{2}{*}{\makecell[c]{Number of \\Points}} 
    & \multicolumn{3}{c}{Total Time (ms)} & \multicolumn{3}{c}{Average Time ($\mu$s)} \\
    \cline{5-7} \cline{8-10}
    & & & & SISL & Parasolid & M-rep (ours) & SISL & Parasolid & M-rep (ours) \\
    \hline
    \multirow{3}{*}{(j)} & \multirow{3}{*}{10} & \multirow{3}{*}{4} & $1 \times 10^{4}$  & 41.8  &  1102  & 3.43  & 4.18  & 110  & 0.343 \\
                             &  &  & $5 \times 10^{4}$ & 156   & 5787 & 5.60  & 3.11  & 116  & 0.112 \\
                             &  &  & $1 \times 10^{5}$  & 307  &  11166  & 6.66  & 3.07  & 112  & 0.067 \\
    \hline
    \multirow{3}{*}{(k)} & \multirow{3}{*}{12} & \multirow{3}{*}{5}& $1 \times 10^{4}$  & 41.3  &  1191  & 5.48  & 4.13  & 119  & 0.548 \\
                             &  &  & $5 \times 10^{4}$ & 204   & 6203 & 7.90  & 4.07  & 124  & 0.158 \\
                             &  &  & $1 \times 10^{5}$  & 406  & 12275  & 11.1  & 4.06  & 123  & 0.111 \\
    \hline
    \multirow{3}{*}{(l)} & \multirow{3}{*}{14} & \multirow{3}{*}{6} & $1 \times 10^{4}$  & 62.0  &  1714  & 13.1  & 6.20  & 171  & 1.31 \\
                             &  &  & $5 \times 10^{4}$ & 244   & 7352 & 18.8  & 4.88  & 147  & 0.377 \\
                             &  &  & $1 \times 10^{5}$  & 494  & 13927  & 24.5  & 4.94  & 139  & 0.245 \\
    \hline
    \multirow{3}{*}{(m)} & \multirow{3}{*}{14} & \multirow{3}{*}{4} & $1 \times 10^{4}$  & 72.5  &  1465  & 5.33  & 7.25  & 147  & 0.533 \\
                             &  &  & $5 \times 10^{4}$ & 288   & 6604 & 8.46  & 5.76  & 132  & 0.169 \\
                             &  &  & $1 \times 10^{5}$  & 587  &  13385  & 11.5  & 5.87  & 134  & 0.115 \\
    \hline
    \multirow{3}{*}{(n)} & \multirow{3}{*}{18} & \multirow{3}{*}{5} & $1 \times 10^{4}$  & 114  &  1750  & 2.34  & 11.4  & 175  & 0.234 \\
                             &  &  & $5 \times 10^{4}$ & 559   & 7735 & 4.61  & 11.2  & 174  & 0.092 \\
                             &  &  & $1 \times 10^{5}$  & 1123  & 14843  & 6.97  & 11.2  & 148  & 0.070 \\
    \hline
    \multirow{3}{*}{(o)} & \multirow{3}{*}{22} & \multirow{3}{*}{6} & $1 \times 10^{4}$  & 182  &  2019  & 10.9  & 18.2  & 202  & 1.09 \\
                             &  &  & $5 \times 10^{4}$ & 892   & 8724 & 17.5  & 17.8  & 174  & 0.349 \\
                             &  &  & $1 \times 10^{5}$  & 1790  & 17083  & 26.7  & 17.9  & 171  & 0.267 \\
    \hline
    \multirow{3}{*}{(p)} & \multirow{3}{*}{35} & \multirow{3}{*}{4} & $1 \times 10^{4}$  & 217  &  3528  & 4.32  & 21.7  & 353  & 0.432 \\
                             &  &  & $5 \times 10^{4}$ & 1060   & 18346 & 13.8  & 21.2  & 367  & 0.277 \\
                             &  &  & $1 \times 10^{5}$  & 2058  & 34186 & 20.5  & 20.6  & 342  & 0.205 \\
    \hline
    \multirow{3}{*}{(q)} & \multirow{3}{*}{46} & \multirow{3}{*}{5} & $1 \times 10^{4}$  & 401  &  2945  & 5.04  & 40.1  & 295  & 0.504 \\
                             &  &  & $5 \times 10^{4}$ & 1980   & 13512 & 18.5  & 39.6  & 270  & 0.371 \\
                             &  &  & $1 \times 10^{5}$  & 3937  & 27627  & 28.1  & 39.4  & 276  & 0.281 \\
    \hline
    \multirow{3}{*}{(r)} & \multirow{3}{*}{57} & \multirow{3}{*}{6} & $1 \times 10^{4}$  & 679  &  3521  & 12.1  & 67.9  & 352  & 1.21 \\
                             &  &  & $5 \times 10^{4}$ & 3320   & 16935 & 23.5  & 66.4  & 339  & 0.469 \\
                             &  &  & $1 \times 10^{5}$  & 6612  & 32267  & 39.4  & 66.1  & 323  & 0.394 \\
    \hline
    \end{tabularx}
    \label{tab: SISL-vs-Parasolid-vs-Mrep_3D}
\end{table*}

\begin{table*}[htb]
    \centering
    \small 
    \renewcommand{\arraystretch}{0.8} 
    \setlength\extrarowheight{2pt} 
    \setlength{\abovecaptionskip}{0cm}
    \caption{Inversion Comparisons of SISL, Parasolid, and M-rep method on 3D cases.}
    \setlength{\tabcolsep}{4mm} 
    \begin{tabularx}{1.0\textwidth}{
        >{\centering\arraybackslash}>{\hsize=.5\hsize\linewidth=\hsize}X
        >{\centering\arraybackslash}>{\hsize=.5\hsize\linewidth=\hsize}X
        >{\centering\arraybackslash}>{\hsize=.5\hsize\linewidth=\hsize}X
        >{\centering\arraybackslash}>{\hsize=1.5\hsize\linewidth=\hsize}X
        >{\centering\arraybackslash}X
        >{\centering\arraybackslash}X
        >{\centering\arraybackslash}X
        >{\centering\arraybackslash}X
        >{\centering\arraybackslash}X
        >{\centering\arraybackslash}X
    }
    \hline
    \multirow{2}{*}{Model} & \multirow{2}{*}{\makecell[c]{Knots \\Length}} & \multirow{2}{*}{Degree} & \multirow{2}{*}{\makecell[c]{Number of \\Points}} 
    & \multicolumn{3}{c}{Total Time (ms)} & \multicolumn{3}{c}{Average Time ($\mu$s)} \\
    \cline{5-7} \cline{8-10}
    & & & & SISL & Parasolid & M-rep (ours) & SISL & Parasolid & M-rep (ours)\\
    \hline
    \multirow{3}{*}{(j)} & \multirow{3}{*}{10} & \multirow{3}{*}{4} & $1 \times 10^{4}$  & 39.5  &  1131  & 4.42  & 3.95  & 113  & 0.442 \\
                             &  &  & $5 \times 10^{4}$ & 151   & 5707 & 5.60  & 3.02  & 114  & 0.112 \\
                             &  &  & $1 \times 10^{5}$  & 299  &  11231  & 6.66  & 2.99  & 112  & 0.067 \\
    \hline
    \multirow{3}{*}{(k)} & \multirow{3}{*}{12} & \multirow{3}{*}{5}& $1 \times 10^{4}$  & 34.1  &  1223  & 6.52  & 3.41  & 122  & 0.652 \\
                             &  &  & $5 \times 10^{4}$ & 158   & 6153 & 8.05  & 3.15  & 123  & 0.161 \\
                             &  &  & $1 \times 10^{5}$  & 305  & 12334  & 13.1  & 3.05  & 123  & 0.131 \\
    \hline
    \multirow{3}{*}{(l)} & \multirow{3}{*}{14} & \multirow{3}{*}{6} & $1 \times 10^{4}$   & 69.0  &  1684  & 12.3  & 6.90  & 168  & 1.23 \\
                             &  &  & $5 \times 10^{4}$ & 343   & 7293 & 17.1  & 6.86  & 146  & 0.341 \\
                             &  &  & $1 \times 10^{5}$  & 704  & 13991  & 24.8  & 7.04  & 140  & 0.248 \\
    \hline
    \multirow{3}{*}{(m)} & \multirow{3}{*}{14} & \multirow{3}{*}{4} & $1 \times 10^{4}$  & 69.2  &  1483  & 6.30  & 6.92  & 148  & 0.630 \\
                             &  &  & $5 \times 10^{4}$ & 346   & 6801 & 8.15  & 6.93  & 136  & 0.163 \\
                             &  &  & $1 \times 10^{5}$  & 691  &  13326  & 11.9  & 6.91  & 133  & 0.119 \\
    \hline
    \multirow{3}{*}{(n)} & \multirow{3}{*}{18} & \multirow{3}{*}{5} & $1 \times 10^{4}$   & 114  &  1759  & 2.34  & 11.4  & 176  & 0.234 \\
                             &  &  & $5 \times 10^{4}$ & 559   & 7513 & 5.96  & 11.2  & 150  & 0.120 \\
                             &  &  & $1 \times 10^{5}$  & 1123  & 14744  & 6.76  & 11.2  & 147  & 0.068 \\
    \hline
    \multirow{3}{*}{(o)} & \multirow{3}{*}{22} & \multirow{3}{*}{6} & $1 \times 10^{4}$  & 184  &  1993  & 8.47  & 18.4  & 200  & 0.847 \\
                             &  &  & $5 \times 10^{4}$ & 876   & 8799 & 11.6  & 17.5  & 176  & 0.232 \\
                             &  &  & $1 \times 10^{5}$  & 1753  & 17021  & 16.1  & 17.5  & 170  & 0.161 \\
    \hline
    \multirow{3}{*}{(p)} & \multirow{3}{*}{35} & \multirow{3}{*}{4} & $1 \times 10^{4}$  & 210  &  3316  & 4.63  & 21.0  & 331  & 0.463 \\
                             &  &  & $5 \times 10^{4}$ & 1027   & 18596 & 13.6  & 20.5  & 372  & 0.271 \\
                             &  &  & $1 \times 10^{5}$  & 2085  & 35113 & 20.1  & 20.9  & 351  & 0.201 \\
    \hline
    \multirow{3}{*}{(q)} & \multirow{3}{*}{46} & \multirow{3}{*}{5} & $1 \times 10^{4}$   & 387  &  2988  & 5.00  & 38.7  & 299  & 0.500 \\
                             &  &  & $5 \times 10^{4}$ & 1973   & 13137 & 15.8  & 39.5  & 263  & 0.316 \\
                             &  &  & $1 \times 10^{4}$  & 3845  & 26924  & 27.3  & 38.5  & 269  & 0.273 \\
    \hline
    \multirow{3}{*}{(r)} & \multirow{3}{*}{57} & \multirow{3}{*}{6} & $1 \times 10^{4}$  & 670  &  3492  & 13.2  & 67.0  & 349  & 1.32 \\
                             &  &  & $5 \times 10^{4}$ & 3264   & 16478 & 23.6  & 65.2  & 330  & 0.471 \\
                             &  &  & $1 \times 10^{5}$  & 6544  & 32219  & 39.2  & 65.4  & 322  & 0.392 \\
    \hline
    \end{tabularx}
    \label{tab: SISL-vs-Parasolid-vs-Mrep_3D_inversion}
\end{table*}

\begin{table}[H]
\centering
\caption{Comparison of analytical and Newton methods for computing roots of $E'(t)$ on GPU.}
\begin{tabular}{|c|c|}
\hline
\textbf{Method} & \textbf{Computation Speed} \\
\hline
Analytical Method& 52us \\
\hline
Newton Method & 214us \\
\hline
\end{tabular}
\label{tab:analytical}
\end{table}

\end{document}